\RequirePackage{fix-cm}
\documentclass[smallextended]{svjour3}       
\smartqed  
\usepackage{graphicx}
\usepackage{color}
\journalname{General Relativity and Gravitation}
\begin{document}

\title{Ho{\v r}ava's quantum gravity illustrated \\
by embedding diagrams of the Kehagias-Sfetsos spacetimes
}

\titlerunning{Embedding diagrams of the Kehagias-Sfetsos spacetimes}        

\author{Kate\v{r}ina~Goluchov\'a     \and
        Konrad~Kulczycki             \and
				Ronaldo~S.~S.~Vieira           \and
				Zden\v{e}k~Stuchl\'{\i}k     \and
			  W{\l}odek~Klu{\'z}niak       \and
				Marek~Abramowicz
}


\institute{Kate\v{r}ina~Goluchov\'a, Zden\v{e}k~Stuchl\'{\i}k, Marek~Abramowicz  \at
              Institute of Physics, Faculty of Philosophy and Science, Silesian University in Opava, \\
							Bezru{\v c}ovo n{\'a}m. 13, CZ-746 01 Opava, Czech  Republic\\
              \email{katka.g@seznam.cz}           
           \and
           Ronaldo~S.~S.~Vieira, W{\l}odek~Klu{\'z}niak, Marek~Abramowicz \at
              Copernicus Astronomical Center, \\
							ul. Bartycka 18, PL-00-716, Warszawa, Poland
						\and
						Ronaldo~S.~S.~Vieira \at
						Instituto de F{\'{\i}}sica ``Gleb Wataghin'',\\
						Universidade Estadual de Campinas, 13083-859, Campinas, SP, Brazil
						\and
						Ronaldo~S.~S.~Vieira \at
						 Instituto de Astronomia, Geof\'{i}sica e Ci\^encias Atmosf\'{e}ricas,\\ 
						 Universidade de S\~{a}o Paulo, 05508-090, S\~{a}o Paulo, SP, Brazil
						\and
						Marek~Abramowicz \at
						Physics Department, Gothenburg University, SE-412-96 G{\"o}teborg, Sweden
}

\date{Received: date / Accepted: date}

\maketitle

\begin{abstract}
  Possible astrophysical consequences of the Ho{\v r}ava quantum gravity theory have been
  recently studied by several authors. They usually employ the Kehagias-Sfetsos (KS) spacetime
  which is a spherically symmetric vacuum solution of a specific version of Ho{\v r}ava's
  gravity. The KS metric has several unusual geometrical properties that in the present
  article we examine by means of the often used technique of {\it embedding diagrams}. We pay
   particular attention to the transition between naked singularity and black-hole states, which
  is possible along some particular sequences of the KS metrics. 
	
\end{abstract}

\section{Introduction}
\label{intro}

%
Black holes almost certainly exist in the real universe. In studying their properties,
astrophysicists usually use two famous exact solutions of Einstein's field
equations of his general relativity theory --- the Schwarzschild solution that describes a
static, spherically symmetric gravity of a non-rotating black hole, and the Kerr solution
which describes a stationary, axially symmetric gravity of a rotating black hole. The
Reissner-Nordstr{\"o}m solution which describes a static, spherically symmetric gravity of a
charged, non-rotating black hole is considered to be astrophysically unrealistic, because
charged macroscopic objects quickly neutralize by accreting plasma of the opposite sign. The
Schwarzschild and Kerr black holes are now quite familiar objects in astrophysics.

However, black holes in the real universe may be different from those
of the Schwarzschild or of the Kerr type. First, Einstein's theory of
gravitation has not yet been tested in its strong field limit relevant
to black holes. A variety of alternative gravity theories were studied and 
theoretically and observationally tested, in the strong gravity limit. Firstly,
there is a number of papers related to
braneworld black holes \cite{dad-etal:2000:PLB:,ali+gun:2005:PRD:,Sche-Stu:2013:JCAP:} or stringy black holes \cite{gib+mae:1988:NPB:,sen:1992:PRL:,has+sen:1992:NP:,bla+bla:2001:CQG:,hio+miy:2008:PRD:}. On the other hand, there is an extensive study 
of regular black holes where general relativity is combined with non-linear electrodynamics \cite{bar:1968:,ayo+gar:1998:PRL:,ayo:1999:PLB:,ayo+gar:1999:GRG:,ayo+gar:2000:PLB:,sche+stu:2015:IJMP:,sche+stu:2015:JCAP:,gho+man:2015:EPJ:}. {Inspired by some ideas of quantum theory, static and rotating, regular
(i.e. singularity-free), black hole spacetimes have been proposed \cite{dym:1992:GRG:,mod:2004:PRD:,nic-etal:2006:PLB:,bam:2013:PRD:,li+bam:2013:PRD:,liu-etal:2014:PRD:,bam-etal:2014:EPJC:,dym+gal:2015:CQG:}, and also their behavior due to gravitational collapse has been studied \cite{zha-etal:2015}. The solution representing regular black holes in the de Sitter universe has
been found in \cite{mat-etal:2008:MPLA:}}. Alternatively, string 
theory could allow for existence of superspinars \cite{gim+hor:2009:PLB:,stu+schee:2013:CQG:}. Secondly, although the quantum gravity theory is still
unknown, there are numerous hints that quantum effects may influence
strong gravity. Several attempts have been made to describe this
influence. Ho{\v r}ava's theory \cite{Hor:2009:PHYSR4:,Hor:2009:PHYSRL:}
 is a particular recent example. Like
some other quantum gravity theories, it points to interesting
possibilities that may be directly relevant to astrophysics: (a) the
existence of naked singularities with observational consequences that
may be calculated, (b) a dynamical phenomenon of 
``antigravity''~\cite{Vieira:2014:PHYSR4:}. It
seems that these two features are genuine for a wide class of
Einsteinian (e.g., Reissner-Nordstr{\"o}m) and non-Einsteinian (e.g.,
quantum) strong gravitational fields. In this paper we discuss these
phenomena specifically for the Ho{\v r}ava gravity, in terms of
embedding diagrams. The embedding diagrams help to build intuition
about strongly curved spacetime geometry, which is equivalent to
strong gravity.
%
%
\section{Ho{\v r}ava's quantum gravity}
%
Ho\v{r}ava \cite{Hor:2009:PHYSR4:,Hor:2009:PHYSRL:} proposed a new approach to quantum gravity that is a field theory based on solid state physics ideas \cite{lif:1941}, being invariant under Lifshitz scaling ($t \rightarrow \beta^{z}t, x \rightarrow \beta x$), with the dynamical critical exponent $z$ flowing from $z=3$  in the short wavelength UV limit, to the value $z = 1$ in the large wavelength IR limit. The theory ``could therefore serve as a UV completion of Einstein's general relativity'' \cite{Hor:2009:PHYSR4:}. Gravity is  an emergent property of the IR limit of the theory; thus Newton's constant is given by a constant of the theory $\kappa$, through $ G = \kappa^{2}/32\pi c$, $c$ is being the speed of light. In this field-theoretical version of
quantum gravity the field equations are derived from a Lagrangian that guarantees
Lorentz invariance (and no preferred time) at low energies, but no Lorentz invariance (and a
preferred time) at high energies. Because of different scaling between time and space, the Ho\v{r}ava quantum gravity action at UV limit is not invariant under general relativistic diffeomorphisms, but it has to be invariant under a diffeomorphism preserving foliations: $t \rightarrow t_{\ast}(t)$, $x^{i}\rightarrow x^{i}_{\ast}(t,x^{i})$. The space and time are treated differently in Ho\v{r}ava gravity, then Lorentz violation occurs at all scales at the UV regime, but vanishes in the IR general relativistic limit.

Soon after, Kehagias and Sfetsos \cite{Keh-Sfe:2009:PhysLetB:} derived and
solved a modified version of Ho{\v r}ava's field equations for the spherically symmetric,
static (i.e. a ``Schwarzschild-like'') case. We will call it the KS spacetime (or the KS
metric) for short. Writing 
\begin{equation}
ds^{2} = -N^{2}dt^{2} + g_{ij}(dx^{i} + N^{i}dt)(dx^{j} + N^{j}dt)
\end{equation}
and discarding the cosmological term, the authors of Ref. \cite{Keh-Sfe:2009:PhysLetB:} considered the action of modified Ho\v{r}ava gravity
\begin{eqnarray}
S = &\int dt d^{3} x \sqrt{g}N \{\frac{2}{K^{2}}\left(K_{ij}K^{ij} - \lambda K^{2}\right) - \frac{\kappa^{2}}{2 \nu^{4}}C_{ij}C^{ij} \nonumber\\
& + \frac{\kappa^{2}\mu}{2\nu^{2}}\epsilon^{ijk}R_{il}^{(3)}\nabla_{j}R^{(3)l}_{\,\,\, \,\,\, \,\,\, k} - \frac{\kappa^{2}\mu^{2}}{8}R_{ij}^{(3)}R^{(3)ij}\nonumber\\
& + \frac{\kappa^{2}\mu^{2}}{8(1 - 3\lambda)}\frac{1 - 4\lambda}{4}\left(R^{(3)}\right)^{2} + \mu^{4} R^{(3)}\},
\end{eqnarray}
where $K_{ij} = (\dot{g}_{ij} - \nabla_{i}N_{j} - \nabla_{j}N_{i})/(2N)$ is the extrinsic curvature of the $t=const$ hypersurfaces and $C^{ij}= \epsilon^{ikm}\nabla_{k}(R^{(3)j}_{\,\,\, \,\,\, \,\,\, m} - R^{(3)}\delta_{m}^{j}/4)$ is the Cotton tensor, $R^{(3)}$ is the 3-D Ricci scalar of the hypersurfaces. The constants $\kappa, \nu$ and $\lambda$ are dimensionless, while the constant $\mu$ has mass dimension 1. The terms involving $\nu$ vanish in the spherically symmetric case \cite{Keh-Sfe:2009:PhysLetB:}. 

In this paper we discuss some extreme geometrical properties of the KS
spacetimes that may be relevant for astrophysics. While Ho{\v r}ava's theory itself is not
generally covariant, our calculations in the KS metric are --- they fulfill the standard
rules of a covariant theory familiar from Einstein general relativity. The KS spherically symmetric solution is asymptotically flat in the $\lambda = 1$ case, with $N_{i} = 0$, and $N^{2} = 1/{|g_{rr}|} \equiv f$, where $f$ is the so called lapse function of the spacetime. The obtained solution contains the parameter $\omega = 16\mu^{2}/\kappa^{2} > 0$, in addition to the gravitational mass $M$, and tends to the Schwarzschild solution in the limit of the dimensionless parameters product $\omega M^{2} \rightarrow \infty$ (expressed in geometrized units). Presumably, $\omega$ is a universal constant, if the KS solution correctly describes the gravity of spherically symmetric objects. For large masses, $\omega M^{2} > 1/2$, the Kehagias-Sfetsos solution has event horizons and corresponds to a black hole \cite{Keh-Sfe:2009:PhysLetB:}. The current observational (lower) limits on $\omega$ are compatible with the existence of naked singularities at masses $M< 2.6\times 10^{4}\,M_{\odot}$ \cite{Ior+rug:2010:IJMP:,liu-etal:2011:GRG:,har-etal:2011:PRCA:}.

We conduct our discussion in terms of the {\it embedding diagrams} of the KS metric
``equatorial planes''. The KS geometry is spherically symmetric, and therefore the geometry
of all central planes is the same as that of the equatorial plane. This allows us to
consider, with no loss of generality, these particular embedding diagrams as a faithful
representation for the whole KS geometry. A similar approach was used previously, for
different spacetimes, by several authors. For example,  Kristiansson et al.
 \cite{Kri-Son-Abr:1998:GRG:} analyzed
embedding of the Reissner-Nordstr{\"o}m (RN) 3-spaces. This work is particularly relevant in
the present context. We will see later that RN and KS metrics share several interesting
properties that nicely show up in the embeddings\footnote{See also references
\cite{Stu-Hed:2002:APS:,Stu-Hle:1999:,Stu-Hle-Jur:2000:}, and
\cite{Stu-Hle-Tru:2011:CLAQG:}.}.

We use here the standard embeddings of the conventional ``physical'', i.e. a simply
projected 3-D space but also less standard embeddings of a projected {\it and in addition}
conformally re-scaled  ``optical'' 3-D space. The optical geometry reflects directly some
dynamical properties of the particle and photon motion. For example, geodesic
lines in the KS optical 3-D space coincide with the photon trajectories.
%

\section{The Kehagias-Sfetsos spacetime}
%
An important class of static and spherically symmetric spacetimes has a general form,
\begin{equation}
\mathrm{d} s^{2} = -f(r) \mathrm{d} t^{2}+\frac{1}{f(r)} \mathrm{d} r^{2}+r^{2} \mathrm{d} \theta^{2}+r^{2}\sin^{2}\theta \mathrm{d} \phi^{2}.
\label{line.element.schwarzscild}
\end{equation}
The ``simply projected'' 3-D space at a given time $t$, is defined by a $t$$=$\,const
section of the spacetime. In this 3-D space, one may define the ``equatorial plane'' by
$\theta$$=$$\pi/2$. It has the intrinsic metric
\begin{equation}
\mathrm{d} \ell^2 = \frac{1}{f(r)} \mathrm{d} r^{2} + r^{2} \mathrm{d} \phi^{2}.
\label{equatorial-directly}
\end{equation}
In the classic (non-quantum) Einstein gravity, for all the static, vacuum, spherically symmetric
and asymptotically flat spacetimes, the lapse function $f(r)$ must be necessarily given by its
``Schwarzschild'' form,
%
\begin{equation}
f(r) = 1 - \frac{2M}{r},
\label{f-schwarzscild}
\end{equation}
where $M$ is the mass in ``geometrical'' units of length, $c\,$$=\,$$1\,$$=\,$$G$.

Kehagias and Sfetsos \cite{Keh-Sfe:2009:PhysLetB:} found that in the quantum-gravity version of
(\ref{line.element.schwarzscild})-(\ref{f-schwarzscild}) one should only modify the lapse function
$f(r)$,
\begin{equation}\label{metric.coeficient}
f(r)=1+r^{2}\omega\left(1-\sqrt{1+\frac{4M}{\omega r^{3}}}\right).
\end{equation}
Here $\omega$ is a physical constant (a quantum-gravity parameter) of dimension
[cm$^{-2}$]. It is easy to see that one recovers the classic Schwarzschild solution
(\ref{f-schwarzscild}) when one puts, in the KS solution (\ref{metric.coeficient}) for the
$f(r)$ function, $\omega \rightarrow \infty$.

The location of the event horizon is given for a static metric by $g_{tt} \equiv f(r) = 0$,
which for the KS solution yields \cite{Keh-Sfe:2009:PhysLetB:}
\begin{equation} \label{horizons-locations}
\left( \frac{r}{M} - 1 \right)^2 = 1 - \frac{1}{2 \omega M^2}.
\end{equation}
Clearly, there are {\it two} horizons, inner and outer, of the KS black hole spacetimes if
\begin{equation}\label{horizonts}
\omega>\omega_{c}=\frac{1}{2M^{2}}.
\end{equation}
For $\omega < \omega_{c}$ a {\it naked KS singularity} occurs at $r = 0$, where the Ricci
and Riemann tensor components behave as (see \cite{Keh-Sfe:2009:PhysLetB:}),
\begin{equation}
[{\rm Ricci,~Riemann}] \sim \frac{1}{r^{3/2}} \sim \infty.
\end{equation}
Note that the KS metric is not Ricci flat. Instead, one has (for large $\omega$),
\begin{equation}
[{\rm Ricci}] \sim \frac{1}{\omega} + {\cal O}(\frac{1}{\omega^2})\,.
\end{equation}
For more details see (\cite{stu-sche-apt:2014}).
It is convenient to write the metric (\ref{line.element.schwarzscild}) in the form,
\begin{eqnarray}
\mathrm{d} s^2 &=& \mathrm{e}^{-2\Phi(r)}\,\mathrm{d} s^2_*, \qquad \Phi(r) = - \frac{1}{2}\,\ln f(r), \label{conformal-metric} \\
\mathrm{d} s^2_* &=& -\mathrm{d} t^{2}+\frac{1}{[f(r)]^2}\mathrm{d} r^{2}+ \frac{r^{2}}{f(r)}\mathrm{d}\theta^{2} +
\frac{r^{2}\sin^{2}\theta}{f(r)} \mathrm{d}\phi^{2}.
\label{line.element.optical}
\end{eqnarray}
The ultra-static\footnote{The ultra-static metrics are defined by $\partial_t g_{ik} =0$,
$g_{ti}=0$ and $g_{tt}=-1$. Their properties have been reviewed by Sonego \cite{Sonego:2010:JMP:}.}
metric (\ref{line.element.optical}), which is conformal to the original metric
(\ref{line.element.schwarzscild}) is often called the ``optical geometry'' metric. The
reason for this name follows from the fact that light trajectories are not only (null)
geodesics in the four-dimensional spacetime, but they are also space-like
geodesics in the 3-D space of the
optical geometry (\ref{line.element.optical}). This property is useful in studying the
motion of particles and photons, as often non-trivial information may be easily deduced directly
from the geometry, in particular the geometry of embedding diagrams. This is why the optical
geometry embeddings are helpful in building physical intuition and therefore worth studying.
From (\ref{line.element.optical}) one deduces that the line element on the 2-D equatorial
plane in the KS optical space is given by
\begin{equation}
\mathrm{d}\ell^2 = \frac{1}{[f(r)]^2}\mathrm{d} r^{2} + \frac{r^{2}}{f(r)}\mathrm{d}\phi^{2}.
\label{equatorial-optical}
\end{equation}
%
%

\section{The embedding procedure}
\label{embedding-procedure}
%
%
Note that the line element on the 2-D equatorial plane in both the simply projected space
(\ref{equatorial-directly}) and the optical space (\ref{equatorial-optical}), may be put in
the form
\begin{eqnarray}
\mathrm{d}\ell^2 &=& g_{rr}(r)\,\mathrm{d} r^2 + g_{\phi\phi}(r)\,\mathrm{d}\phi^2 \nonumber \\
&=& \mathrm{d} r_*^{2} + {\tilde r}^2\,\mathrm{d}\phi^{2},
\label{equatorial-general}
\end{eqnarray}
where $r_*$ is the ``geodesic'' and $\tilde r$ the ``circumferential'' radius of an
$r$\,$=$\,const circle. These two radii are defined by
\begin{equation}
r_*(r) = \int \sqrt{g_{rr}}\,\mathrm{d} r, \qquad {\tilde r}(r) = \sqrt{g_{\phi\phi}}.
\label{geodesic-circumferential}
\end{equation}
Thus, in the simply projected metric we have
\begin{equation}
r_*(r) = \int {\frac{1}{[f(r)]^{1/2}}} \,\mathrm{d} r, \qquad {\tilde r}(r) = r,
\label{geodesic-circumferential-direct}
\end{equation}
and in the optical geometry we have
\begin{equation}
r_*(r) = \int {\frac{1}{f(r)}} \,\mathrm{d} r, \qquad {\tilde r}(r) = \frac{r}{\sqrt{f(r)}}.
\label{geodesic-circumferential-optical}
\end{equation}
The curvature ${\cal K}$ of an $r$\,$=$\,const circle and the Gaussian curvature of the
equatorial plane at the circle are given by
\begin{equation}
{\cal K} = + \frac{1}{\tilde r}\left( \frac{\mathrm{d}{\tilde r}}{\,\mathrm{d} r_*} \right), \qquad
{\cal G} = - \frac{1}{\tilde r}\left( \frac{\mathrm{d}^2{\tilde r}}{\mathrm{d} r_*^2} \right).
\label{Curvature-Gauss}
\end{equation}
The curvature radius of the circle is defined as ${\cal R}=1/{\cal K}$. The Gaussian
curvature ${\cal G}$ has the dimension of [cm$^{-2}$].
%

In the embedding procedure, one determines a function
\begin{equation}
Z = Z(R),
\label{embedding-function}
\end{equation}
such that in the 3-D Euclidean space
\begin{equation}
\mathrm{d}\ell^2 = \mathrm{d} R^2 + \mathrm{d} Z^2 + R^2\,\mathrm{d}\phi^2
\label{euclidean-space}
\end{equation}
the 2-D surfaces $Z = Z(R)$ have the same intrinsic geometry,
\begin{equation}
\mathrm{d}\ell^2 = \left[1 + \left(\frac{\mathrm{d} Z}{\mathrm{d} R}\right)^2 \right]\mathrm{d} R^2 + R^2\,\mathrm{d}\phi^2,
\label{embedding-metric}
\end{equation}
as the intrinsic geometry on the equatorial plane
\begin{equation}
\mathrm{d}\ell^2 = \mathrm{d} r_*^{2} + {\tilde r}^2\,\mathrm{d}\phi^{2}.
\label{equatorial-general-01}
\end{equation}
Equating expressions ($\ref{embedding-metric}$) and ($\ref{equatorial-general-01}$)
one gets equations which determine the embedding
function in the parametric form $Z=Z(R(r))$, with $r$ being the parameter,
\begin{equation}
\left[1 + \left(\frac{\mathrm{d} Z}{\mathrm{d} R}\right)^2 \right]^{1/2}\mathrm{d} R = \mathrm{d} r_*(r), \qquad R = {\tilde r}(r).
\label{embedding-equations-00}
\end{equation}
The explicit solution for $\mathrm{d} Z(r)/\mathrm{d} r$ follows after short algebra,
\begin{eqnarray}
\frac{\mathrm{d} Z}{\mathrm{d} r} &=& \frac{\mathrm{d} R}{\mathrm{d} r}\sqrt{\left(\frac{\mathrm{d} r_*}{\mathrm{d} r}\right)^2\left(\frac{\mathrm{d}{\tilde
r}}{\mathrm{d} r}\right)^{-2} - 1},
\label{dZ-dR} \\
\frac{\mathrm{d} R}{\mathrm{d} r} &=&  \frac{\mathrm{d} {\tilde r}}{\mathrm{d} r}.
\label{dR-dr}
\end{eqnarray}
The functions $r_*(r)$ and ${\tilde r}(r)$ are given in terms of $g_{rr}$ and $g_{\phi\phi}$
by Eq. (\ref{geodesic-circumferential}), and $g_{rr}(r)$ and $g_{\phi\phi}(r)$ are
given by Eqs. (\ref{equatorial-directly}), (\ref{equatorial-optical}) in terms of the
function $f(r)$, which is itself given by (\ref{metric.coeficient}). Knowledge of these
functions allows one to directly integrate (\ref{dZ-dR}) numerically and obtain the
embedding profile $Z(R)$.

Eqs. (\ref{dZ-dR})-(\ref{dR-dr}) do not always determine {\it directly} a function
$Z(R)$, because the relation $F(R,Z)$ is multi-valued and does not determine a unique
mapping $R \rightarrow Z$. This is never a problem in practice, however, as it is always
obvious which specific map should be chosen. The specific choice defines uniquely the
relation, indeed a {\it function}, $Z(R)$. It is this function which we have in mind here.
This remark concerns mostly the region between the ``turning points'' in the optical space
embedding (which we discuss later).

Note that (\ref{dZ-dR}) may be also written as
\begin{equation}\label{condition-embedd-01}
\frac{\mathrm{d} Z}{\mathrm{d} R} = \frac{1}{{\tilde r}}\sqrt{{\cal R}^2 - {\tilde r}^2} = \frac{1}{{\tilde
r}}\sqrt{[{\cal R} - {\tilde r}][{\cal R} + {\tilde r}]}\,.
\end{equation}
Because ${\cal R}$ and ${\tilde r}$ are (by definition) non-negative, the condition that the
function under the square root in Eq. (\ref{condition-embedd-01}) be non-negative
reads
\begin{equation}\label{condition-embedd-02}
{\cal R} \ge {\tilde r}.
\end{equation}
If condition (\ref{condition-embedd-02}) is not fulfilled,
then the embedding in the 3-D
Euclidean space is not possible. 
One knows that  ${\rm sign}({\cal G}) = -{\rm sign}(\mathrm{d}^2{\tilde r}/\mathrm{d} r^2)$.
 It follows that the Gaussian
curvature of the embedded surfaces changes sign as $Z$ increases, both in
simply projected and optical spaces, being always positive in the anti-gravity
region. This change of sign does not depend on the existence of photon
orbits, since it appears for all values of $\omega$ (see for instance Fig. 2
 for the
simply projected case and Figs. 5, 6, 7 for the optical case).
%
%
\section{Embedding of the KS equatorial plane}
\label{section-embedding-KS-plane}
%
For simplicity of notation, we will use from now on a re-scaled (dimensionless) version of
the quantum parameter $\omega$. The re-scaling is defined by $\omega M^2 \rightarrow
\omega$. Similarly, we re-scale length, $x = \{r, R, {\tilde r}, Z, ...\}$, by $x
\rightarrow x/M$. There is no danger of confusion between the two versions of $\omega$ and $x$, so
we will use the same symbols for both of them. For example, the critical value of $\omega$
defined by (\ref{horizonts}) may be simply written as $\omega_c = 1/2$.
%
\subsection{The case of a simply projected space, Eq.~(\ref{equatorial-directly})}
\begin{figure}[htbp]
\begin{center}
  \begin{tabular}{cc}
       \includegraphics[clip,height=2.5cm]{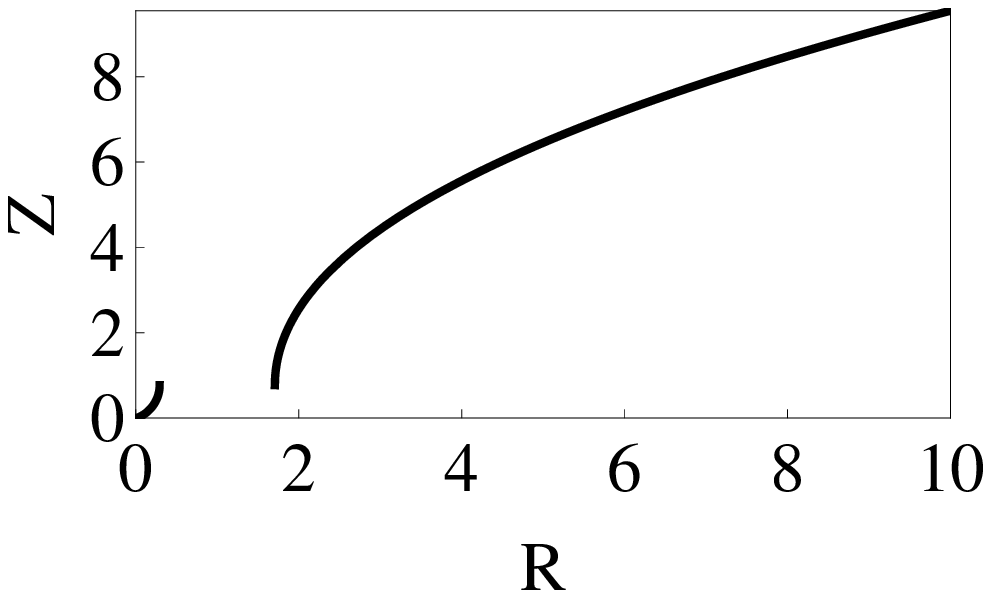}&
        \includegraphics[clip,height=2.5cm]{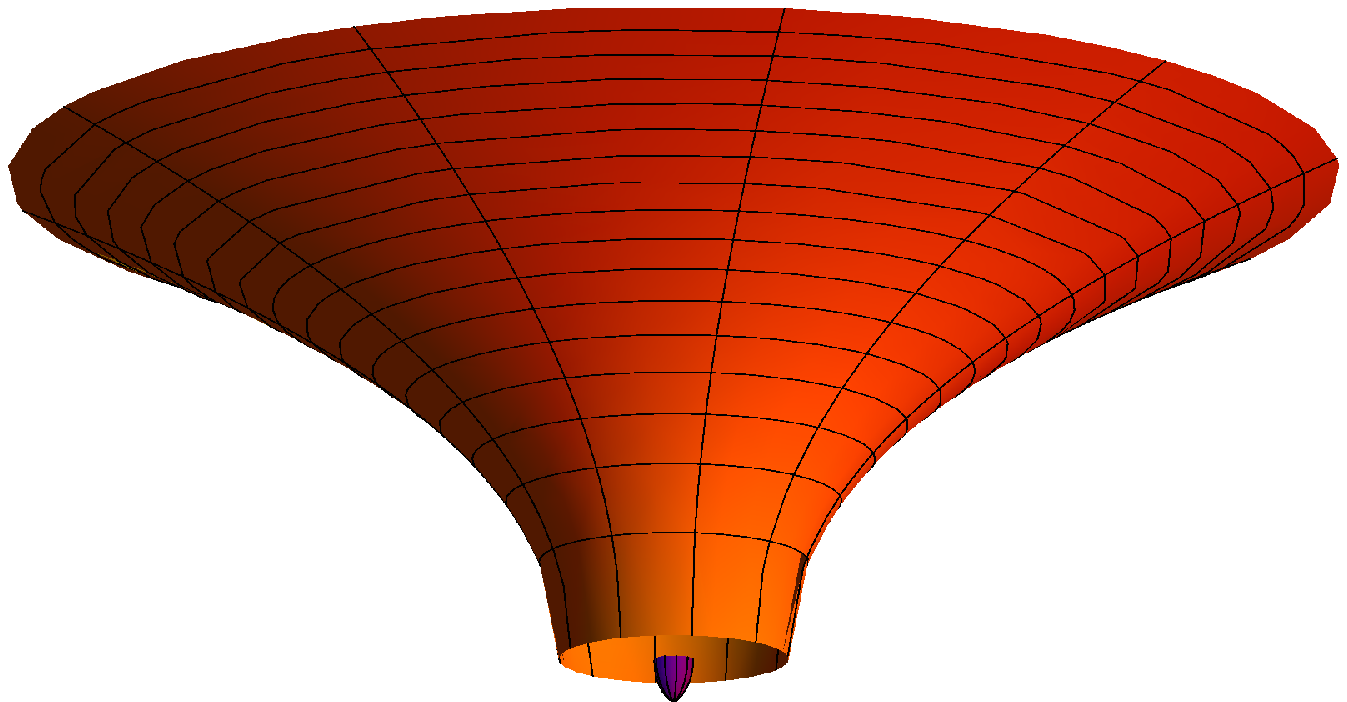}\\
                \includegraphics[clip,height=2.5cm]{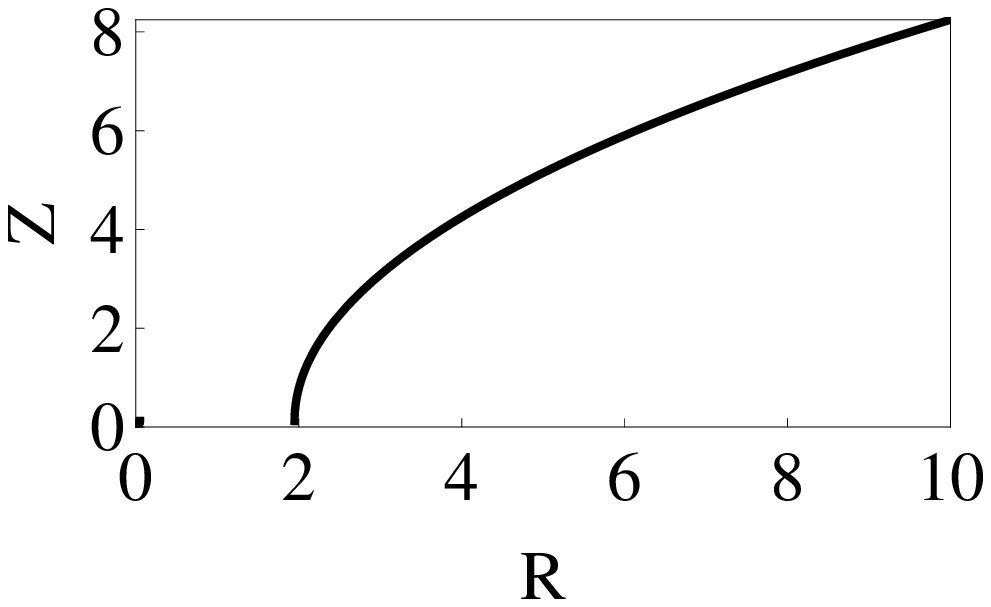}&
        \includegraphics[clip,height=2.5cm]{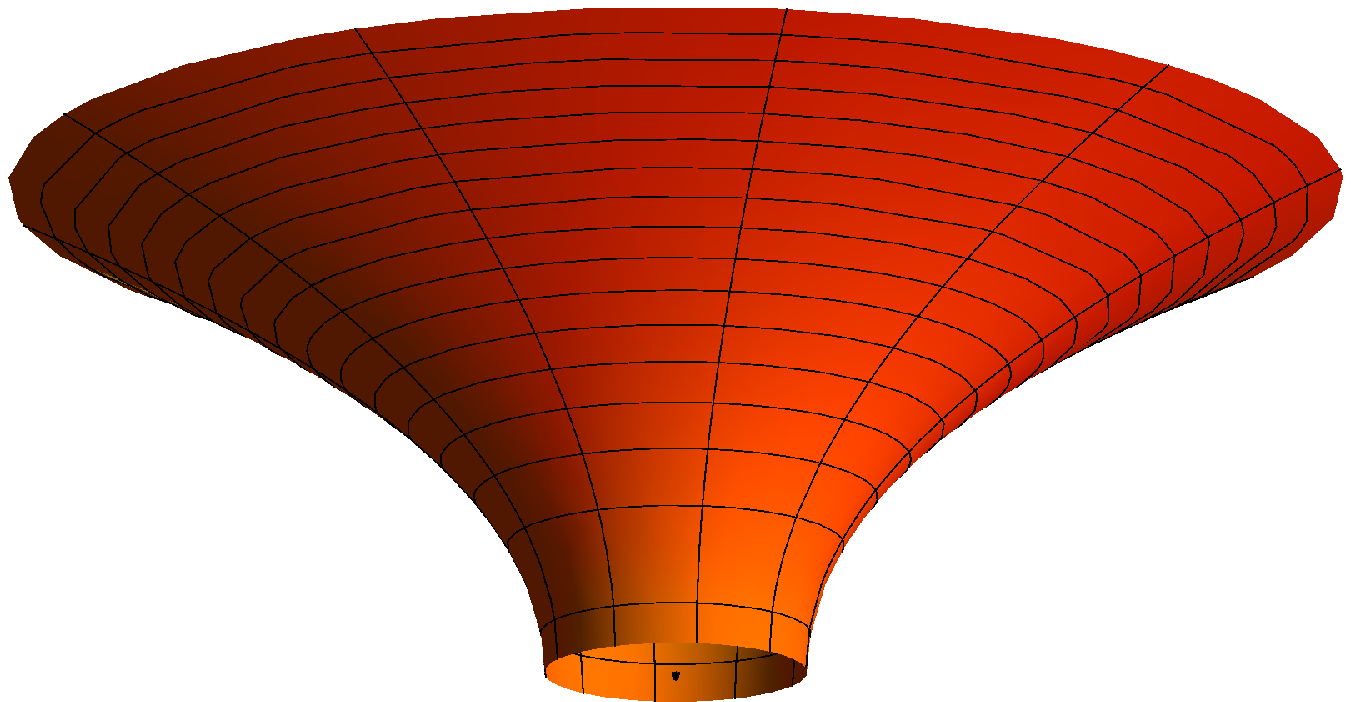}\\
                \end{tabular}
    \caption{Embedding diagrams for the equatorial plane in the simply projected
    3-D space of the KS {\it black hole} spacetime. The left column shows the
    function $Z(R)$, the right column shows the 2-D curved equatorial
    plane embedded into the 3-D Euclidean space. Top diagrams are constructed for $\omega = 1$
    and the bottom ones for $\omega = 5$. The ``outer'' part of the equatorial plane
    ends at the location of the outer horizon. There is also the ``inner'' part
    of the equatorial plane, corresponding to the 3-D space inside the inner horizon. The embedded antigravity region is painted dark violet.
    As the circumferential radii of the outer and inner horizons are different, there is
    a clearly visible discontinuity in geometry. }
    \label{fig_1}
        \end{center}
\end{figure}

%
By using the procedure described in the previous Section, we derive for the function $Z =
Z(r)$ that determines the shape of the embedding profile,
\begin{equation}
\frac{ \mathrm{d} Z}{ \mathrm{d} r} = \sqrt{\frac {-r^{2}\omega\left(1 - \sqrt{1 + \frac{4}{\omega
r^{3}}}\,\right)}{1 + r^{2}\omega\left(1 - \sqrt{1 + \frac{4}{\omega r^{3}}}\,\right)}},
\qquad r = R.
\end{equation}
In the case of the black hole, i.e. $\omega > 1/2$, the whole region $r_+ < r < \infty$
 can be embedded (from the outer horizon to infinity), as the condition
(\ref{condition-embedd-02}) is fulfilled everywhere.
 The region $0<r<r_-$ can also be embedded.
We show examples of this embedding in
Fig. \ref{fig_1}.

\begin{figure}[htbp]
\begin{center}
  \begin{tabular}{cc}
        \includegraphics[clip,height=2.5cm]{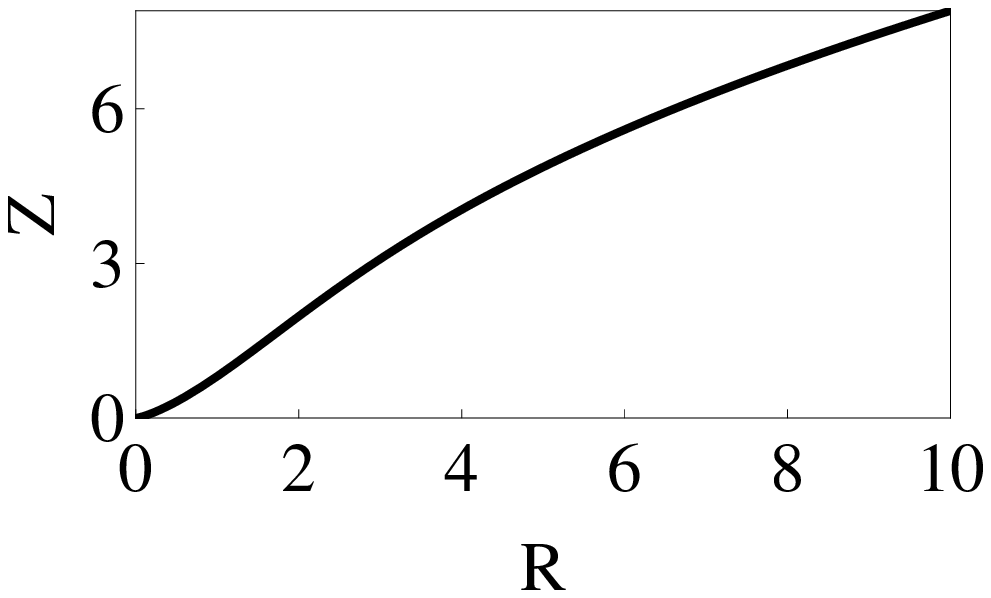}&
        \includegraphics[clip,height=2.5cm]{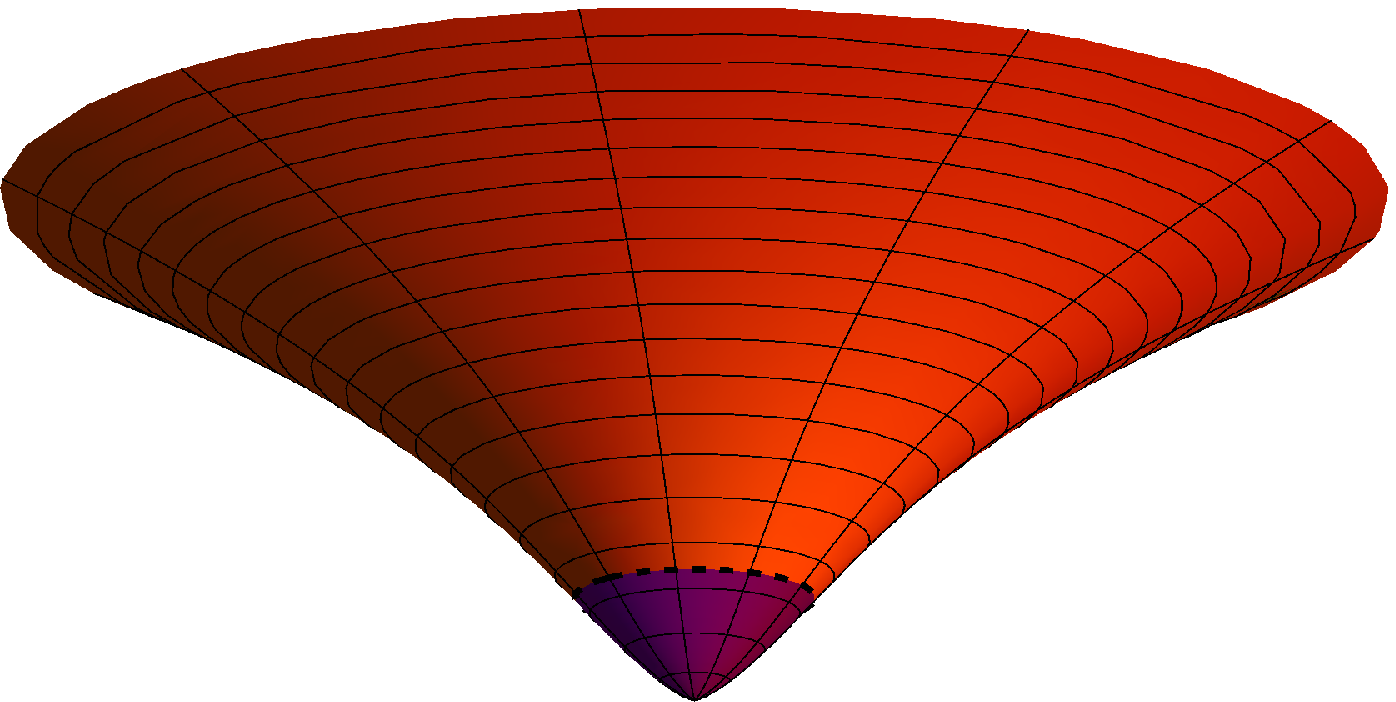}\\
                \includegraphics[clip,height=2.5cm]{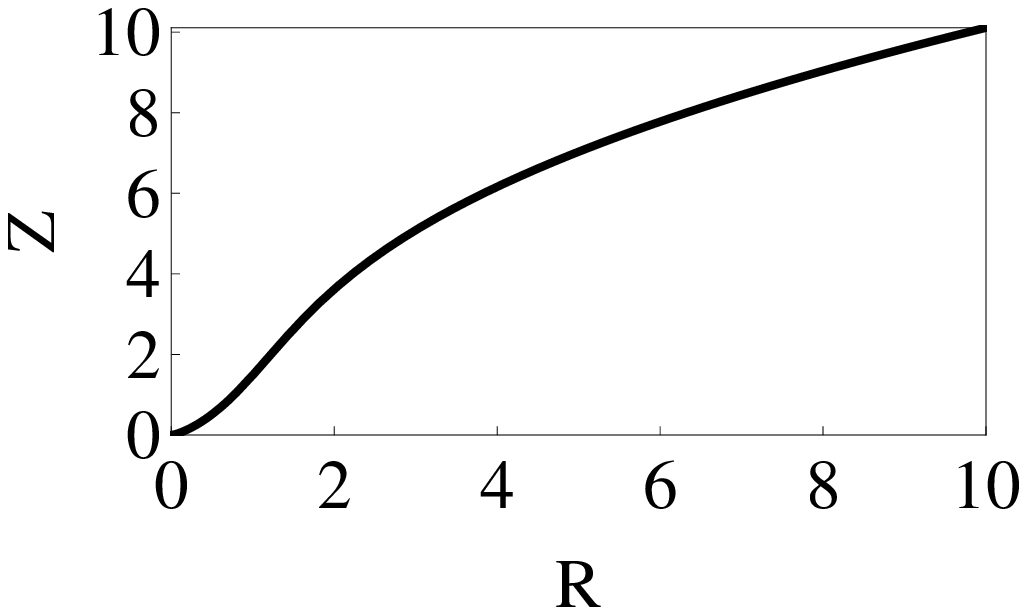}&
        \includegraphics[clip,height=2.5cm]{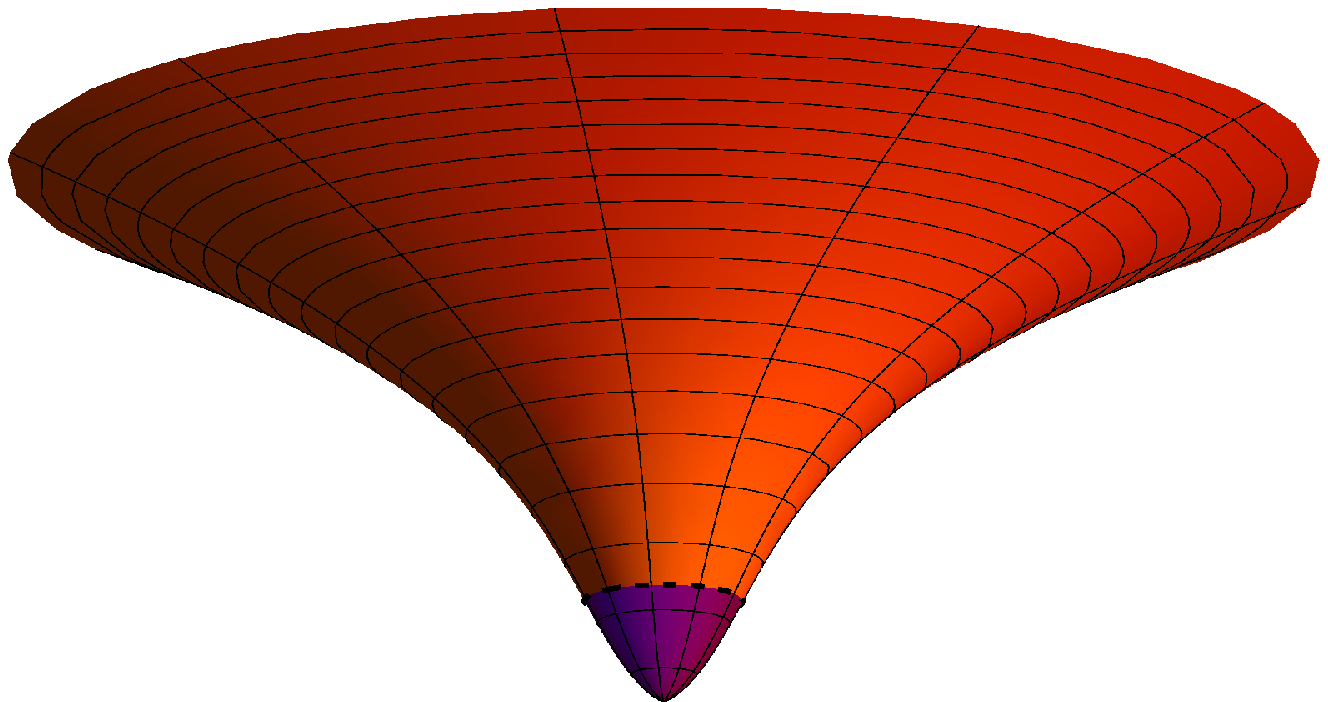}\\
            \end{tabular}
    \caption{Embedding diagrams for the equatorial plane in the simply projected
    3-D space of the KS {\it naked singularity} spacetime.  The top diagram corresponds
    to $\omega = 0.1$ and the bottom one to $\omega = 0.3$. The embedded surfaces go all the way
    to $r=0$, i.e. the location of the naked singularity (the lowest central point on the surfaces). }
    \label{fig_2}
        \end{center}
 \end{figure}

For naked singularities, $\omega < \omega_{c}$, one can construct embedding diagrams for the
entire range of the radial coordinate $0 < r < \infty$. Examples are presented in Fig.
\ref{fig_2}.

\renewcommand{\tabcolsep}{1pt}

%
\subsection{The case of the optical space, Eq.~(\ref{equatorial-optical})}
\label{sub-optical}
%
In this case, the embedding profile is given by equations,
\begin{eqnarray}
\frac{\mathrm{d} Z}{\mathrm{d} r} &=& f^{-1}(r)\sqrt{1 - f^{-1}(r)\left[ 1 - \frac{3}{r}\left(1 +
\frac{4}{\omega r^{3}}\right)^{-\frac{1}{2}}\right]^{2}},
\label{em.form} \\
\frac{\mathrm{d} R}{\mathrm{d} r} &=& f^{-3/2}(r)\left[1 - \frac{3}{r}\left(1 + \frac{4}{\omega
r^{3}}\right)^{-1/2}\right], \\
R &=& \frac{r}{\sqrt{f(r)}}\,.
\end{eqnarray}
The ``turning points'' of the embedding profile are given by the condition
\begin{equation} \label{turning-points}
\frac{\mathrm{d} R}{\mathrm{d} r} = 0.
\end{equation}
According to (\ref{Curvature-Gauss}) they are spacelike
geodesic circles, and therefore they
correspond to the location of circular {\it photon orbits} in the KS spacetime.

From Eq. (\ref{em.form}), we can see that embedding is possible if
\begin{equation} \label{embedding-possible}
1 - f^{-1}(r)\left[ 1 - \frac{3}{r}\left(1 + \frac{4}{\omega
r^{3}}\right)^{-\frac{1}{2}}\right]^{2} \geq 0.
\end{equation}
%
\subsection{Three classes of embedding diagrams \label{three-classes}}
\begin{figure}[htbp]
    \centering
       \includegraphics[width=0.55\linewidth]{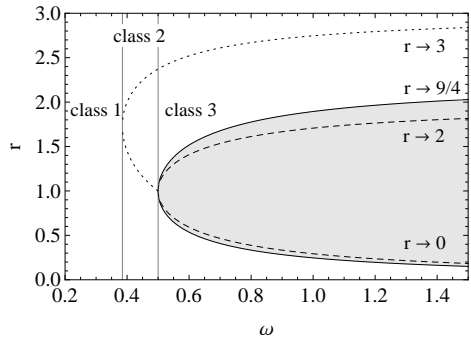}
   \caption{ Relevant radii in KS spacetime.
   The long-dashed line shows location of the horizons. It is given by 
   Eq.~(\ref{horizons-locations}). The short-dashed line shows the location of turning points
in the optical geometry. It is given by
   Eq.~(\ref{turning-points}). The solid line shows limits on embedability (in the optical geometry
   case). It is given by
   Eq.~(\ref{embedding-possible}). The shaded region cannot be embedded. Numbers near the right axis show the asymptotic ($\omega \rightarrow \infty$)
   behavior of the corresponding curves.  The vertexes of the (horizontal) ``parabolas'' that these three lines represent are
   at $(\omega, r) = (\frac{1}{3}\sqrt{\frac{4}{3}}, \sqrt{3})$ and $(\omega, r )= (\frac{1}{2}, 1)$. These two
   locations define the split of the parameter space into three classes, as shown by vertical lines
   and labels, and as described in the text. }
   \label{fig_3}
\end{figure}

%
In terms of $\omega$, the locations of horizons [given by Eq.
(\ref{horizons-locations})], turning points [given by Eq. (\ref{turning-points})], and
the limits on the embedding procedure  [given by Eq.
(\ref{embedding-possible})] are shown in Fig. \ref{fig_3}. The Figure suggests that there
are three classes of embedding diagrams in optical geometry:

{\it class 1}: The range $\omega < \frac{1}{3}\sqrt{ \frac{4}{3}}$ corresponds to naked
singularities. No horizons are present. There are no turning points or embedability limits,
either. The whole region from $r = 0$ to infinity can be embedded. Examples of
embedding
diagrams of this class are pictured in Fig.  \ref{fig_4}.

{\it class 2}: The range $\frac{1}{3}\sqrt{ \frac{4}{3}} \leq \omega < \frac{1}{2}$
corresponds to naked singularities as well. However, turning points exist. Embedding
diagrams for this class are shown in Fig. \ref{fig_5}.

{\it class 3}: The range $\omega > 1/2$. The object is a black hole and there are horizons.
Also, limits on embedability and turning points are present. There is always an
embedability limit outside the outer horizon. Embedding diagrams for this class are shown
in Fig. \ref{fig_6}.
\begin{figure}[htbp]
    \centering
       \includegraphics[width=0.55\linewidth]{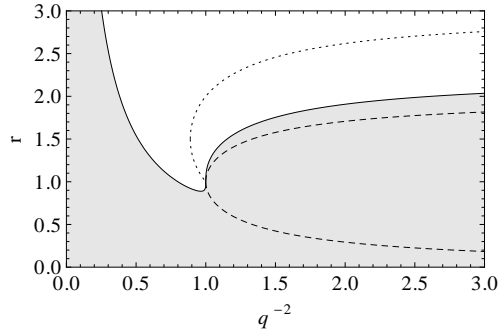}
   \caption{Relevant radii in Reissner-Nordstr\"om spacetime. The lines and the shaded area have the same meaning as in Fig. \ref{fig_3}.
   The radial coordinate is re-scaled as $r\to r/M$ and the charge parameter is $q=Q/M$.
   We see that the radii corresponding to turning points and horizons have the same qualitative behavior as in KS spacetime,
   however the embedability conditions are qualitatively different.}
   \label{fig_3b}
\end{figure}
  \begin{figure}[htbp]
\begin{center}
  \begin{tabular}{ccc}
       \includegraphics[clip,height=2.5cm]{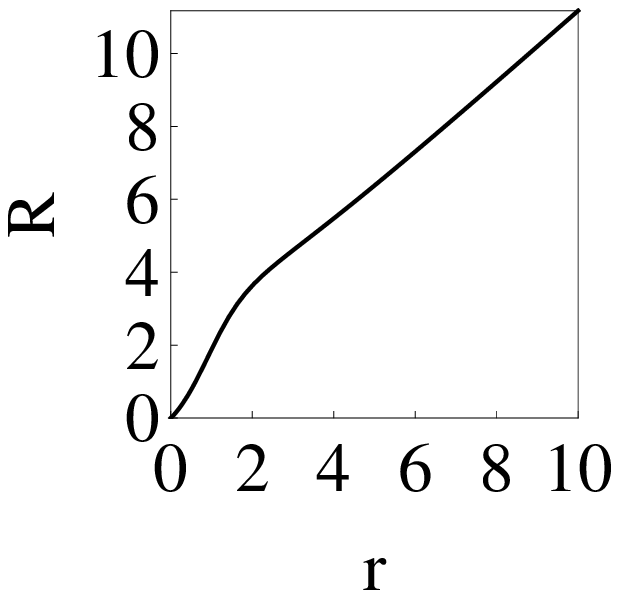}&
        \includegraphics[clip,height=2.5cm]{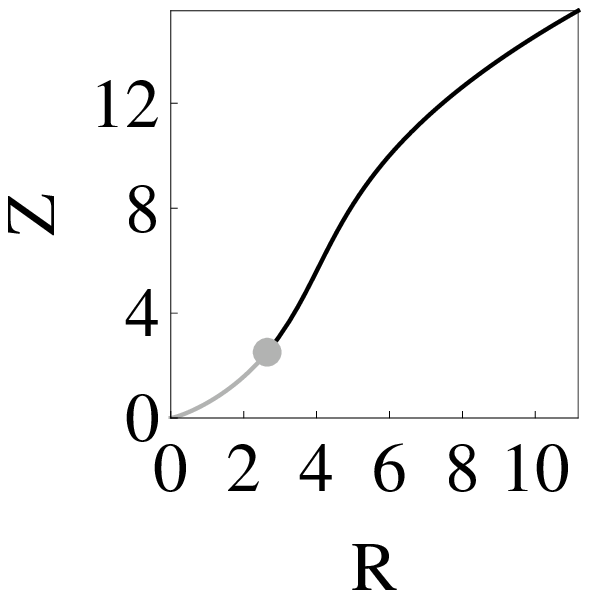}&
                 \includegraphics[clip,height=2.5cm]{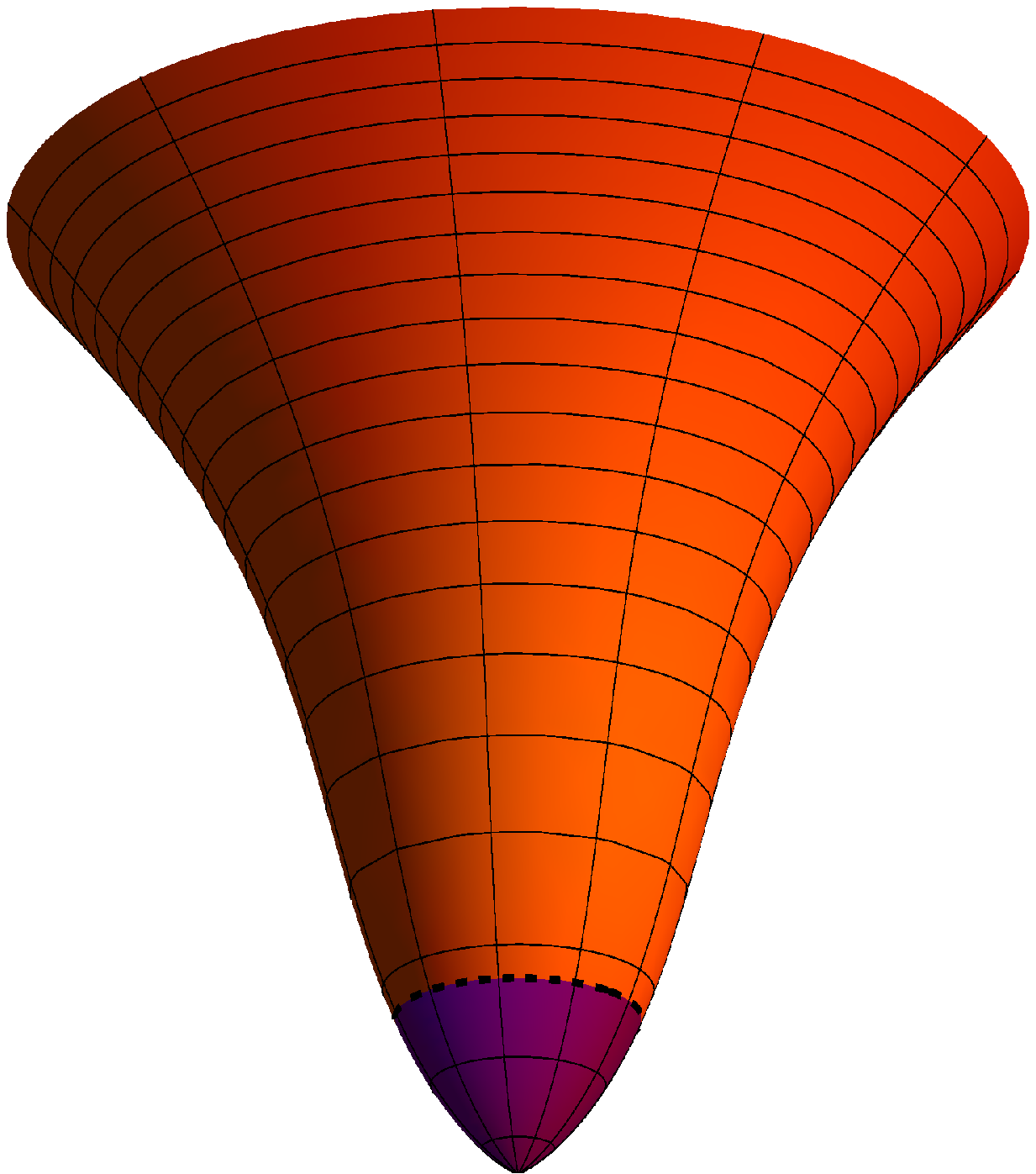}\\
                  \includegraphics[clip,height=2.5cm]{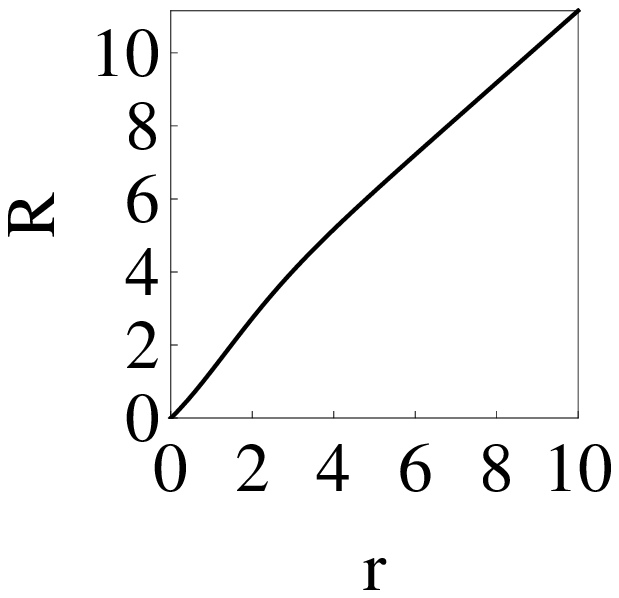}&
        \includegraphics[clip,height=2.5cm]{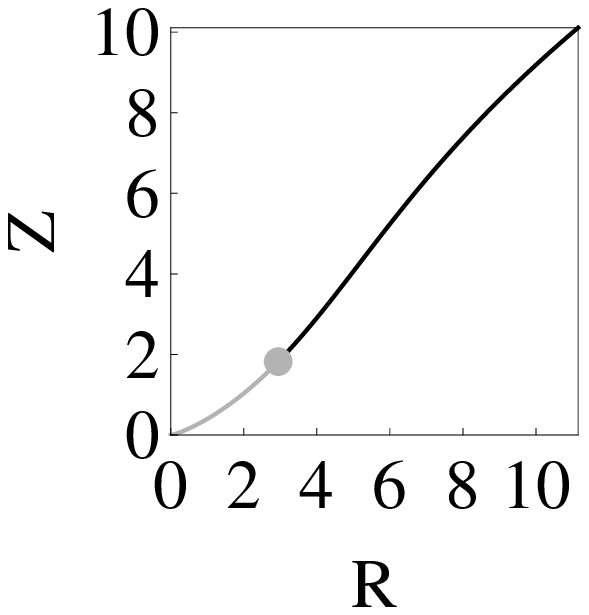}&
                 \includegraphics[clip,height=2.5cm]{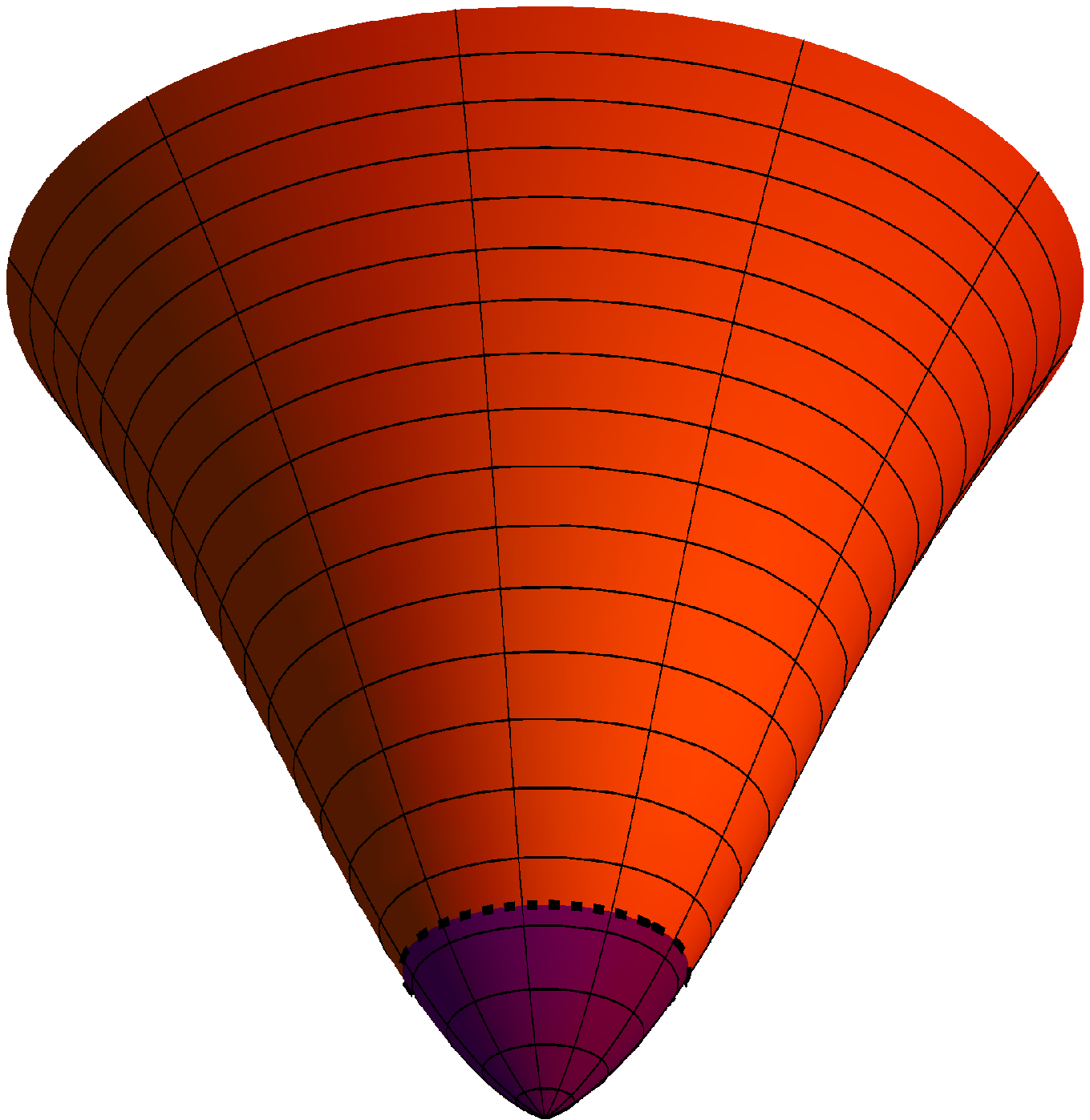}\\
                \end{tabular}
   \caption{Class 1 of the embedding diagrams (a naked singularity). The optical
   geometry case is shown. Top: $\omega = 0.2$. Bottom: $\omega = 0.05$. The left column shows the
   circumferential radius $R(r)$, the middle column the embedding profile $Z(R)$, and the right
   column the embedded surface. The grey color at the profile shows the antigravity region, the
   grey point shows the location of the zero-gravity sphere. The antigravity region is
   indicated by a dark violet color in the embedded surface. See subsection
   \ref{antigravity-centrifugal-reversal} for a more detailed
   explanation. }
    \label{fig_4}
        \end{center}
 \end{figure}
 \begin{figure}[htbp]
\begin{center}
  \begin{tabular}{ccc}
                     \includegraphics[clip,height=2.5cm]{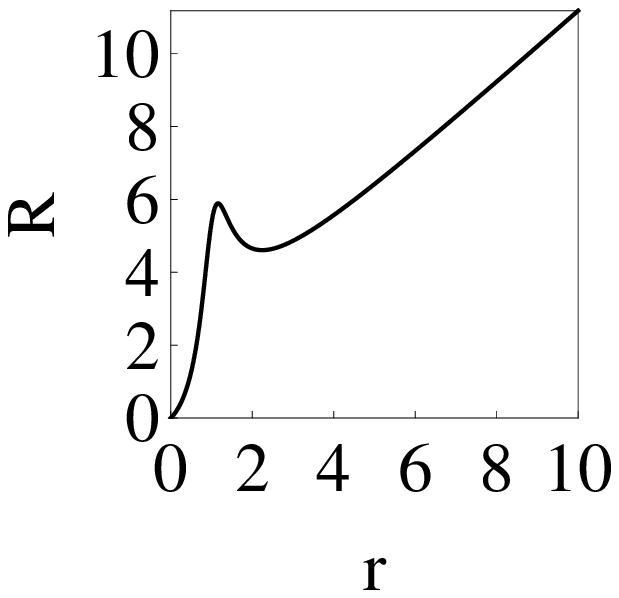}&
        \includegraphics[clip,height=2.5cm]{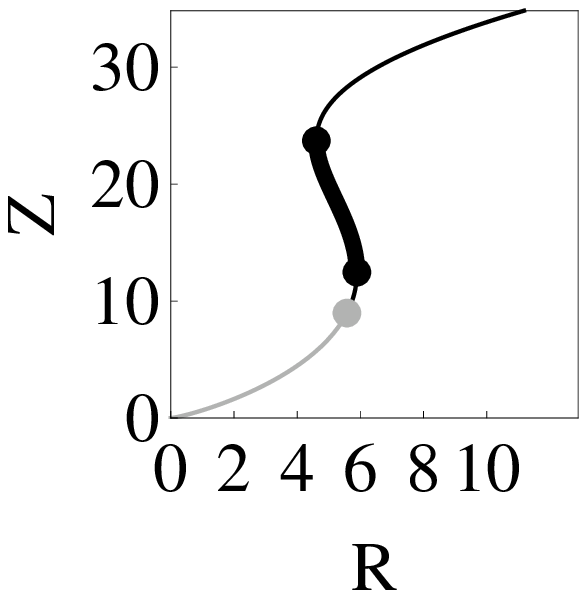}&
                 \includegraphics[clip,height=2.5cm]{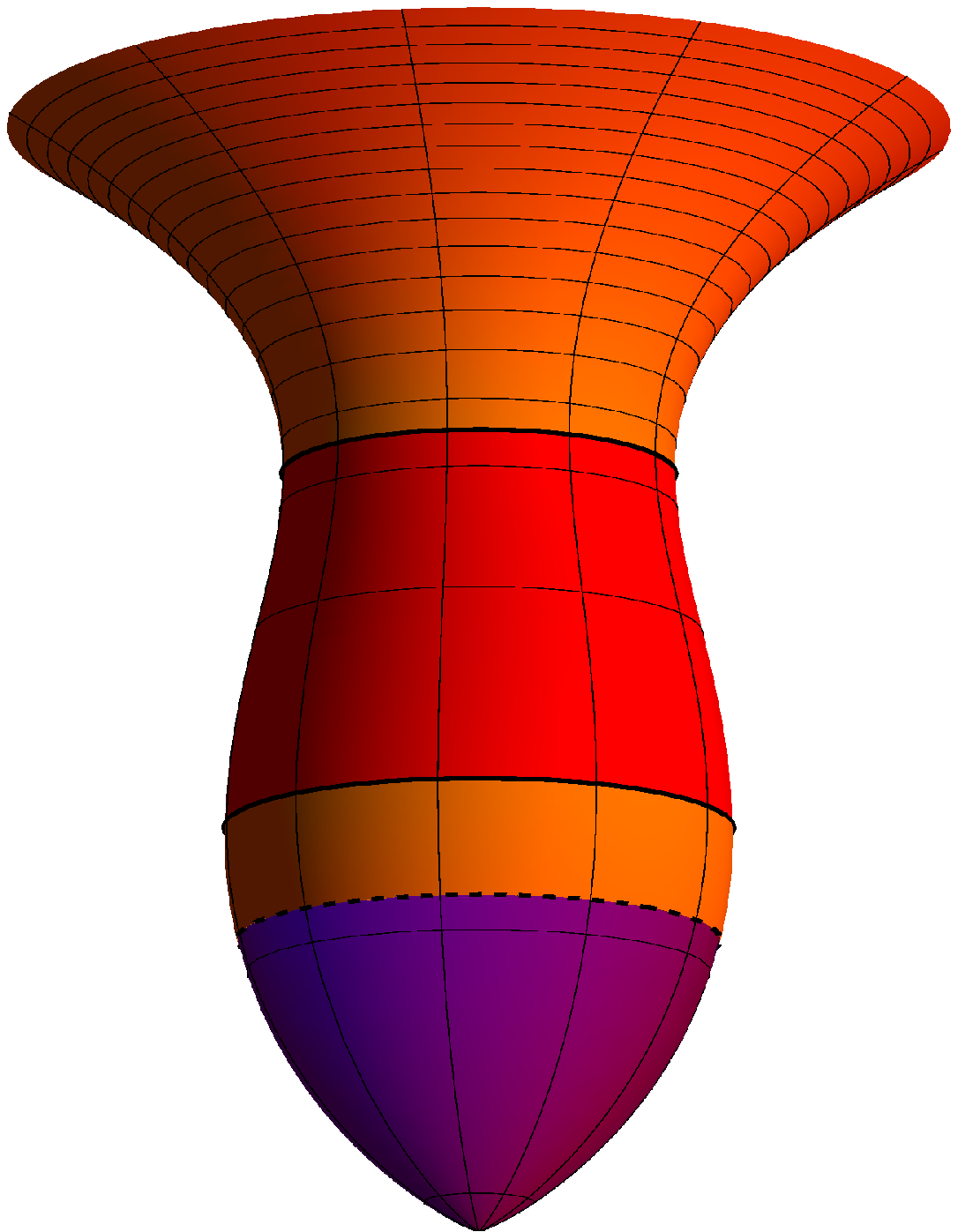}\\
       \includegraphics[clip,height=2.5cm]{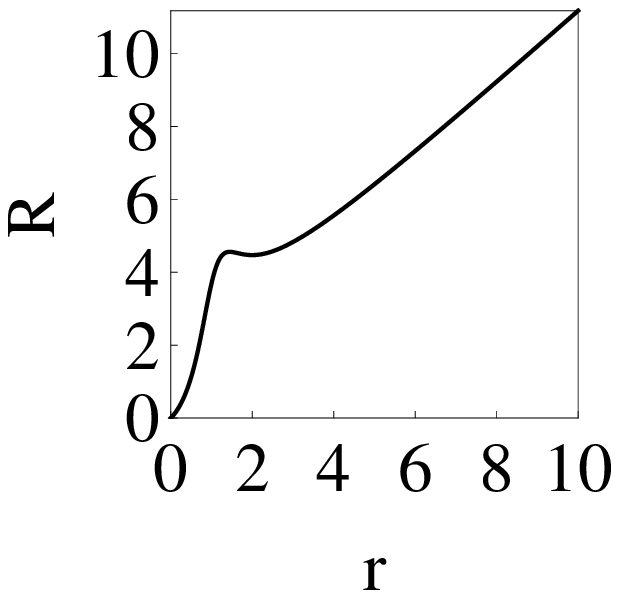}&
        \includegraphics[clip,height=2.5cm]{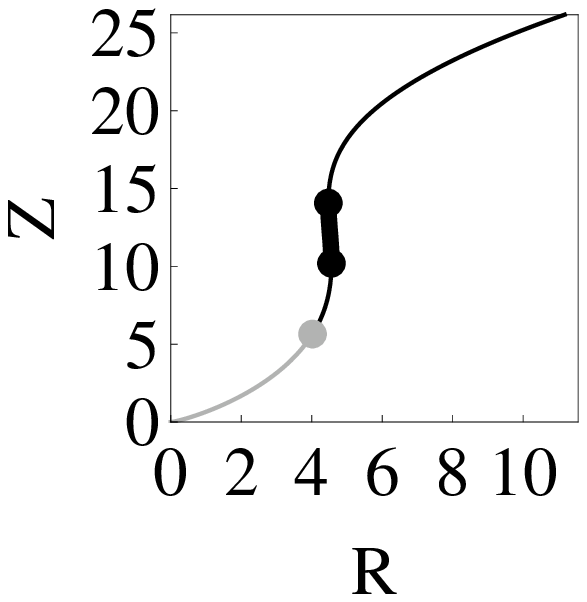}&
                 \includegraphics[clip,height=2.5cm]{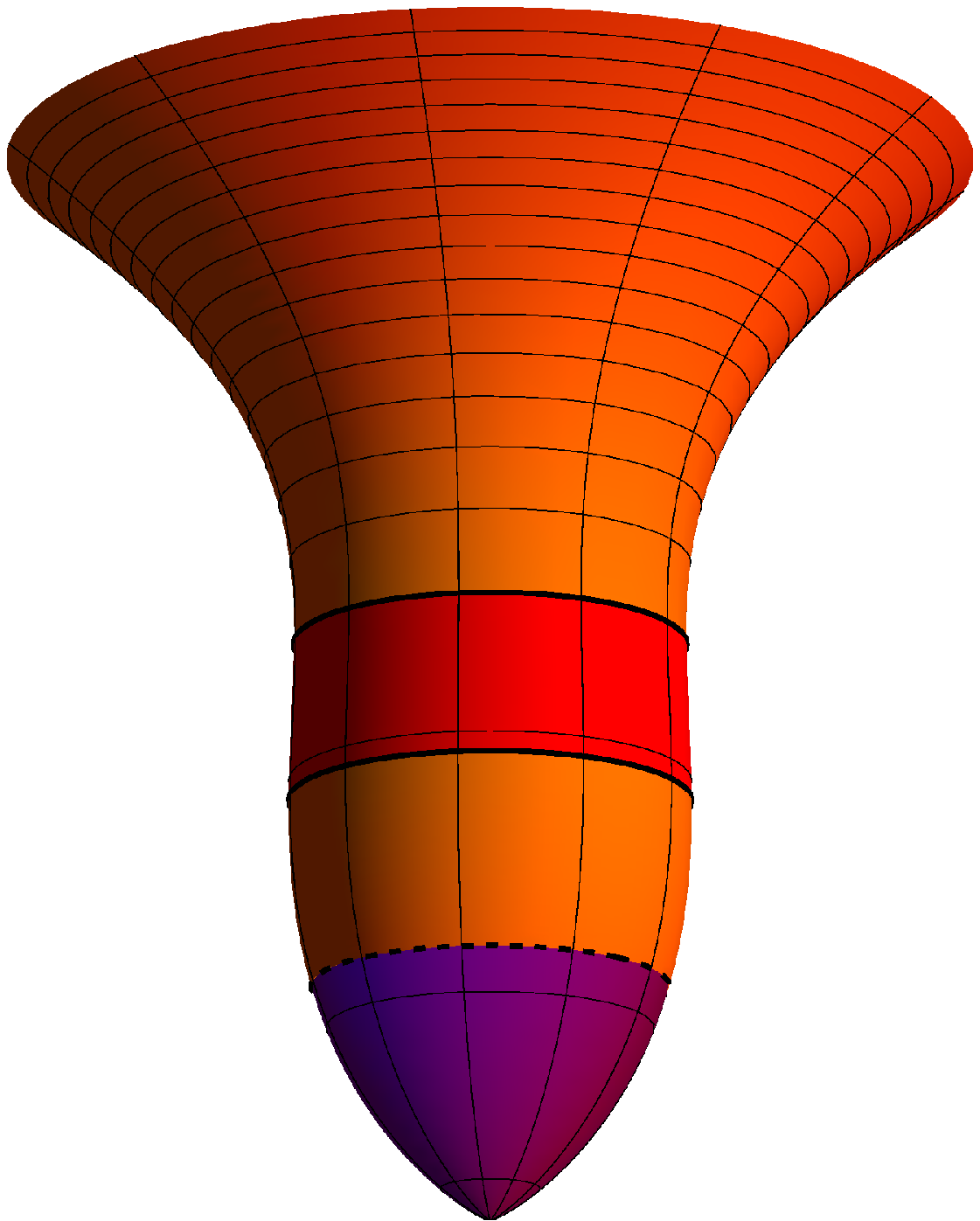}\\
                \end{tabular}
    \caption{Class 2 of the embedding diagrams (naked singularity).
    The optical geometry case is shown. Top: $\omega = 0.45$.
    Bottom: $\omega = 0.4$. The left column shows the
   circumferential radius $R(r)$, the middle column the embedding profile $Z(R)$, and the right
   column the embedded surface. The grey color at the profile shows the antigravity region, the
   large grey dot shows the location of the zero-gravity sphere. 
   The region where centrifugal force is reversed is shown in red.
   The antigravity region is indicated by a dark violet color in the embedded surface.
   The dark large dots show the locations of the
   turning points. They correspond to the circular photon orbits. The photon orbits are indicated
   by solid circles at the embedded surface. See subsection
   \ref{antigravity-centrifugal-reversal} for a more detailed
   explanation.}
   \label{fig_5}
    \end{center}
 \end{figure}
 \begin{figure}[htbp]
\begin{center}

\begin{tabular}{ccc}
                     \includegraphics[clip,height=2.5cm]{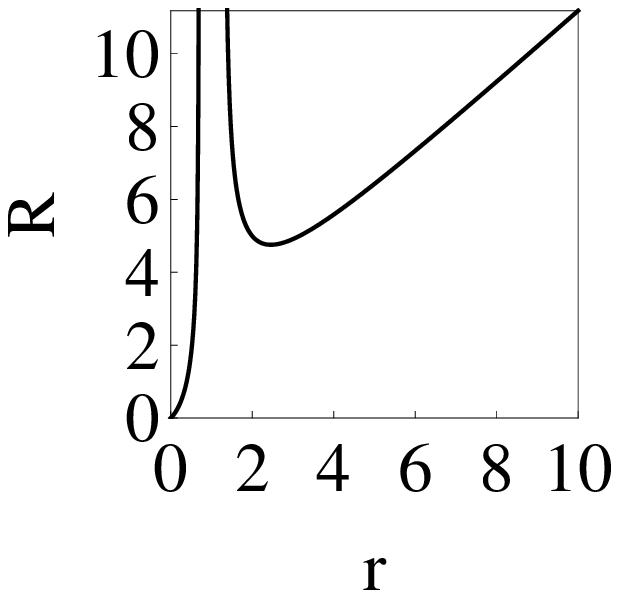}&
        \includegraphics[clip,height=2.5cm]{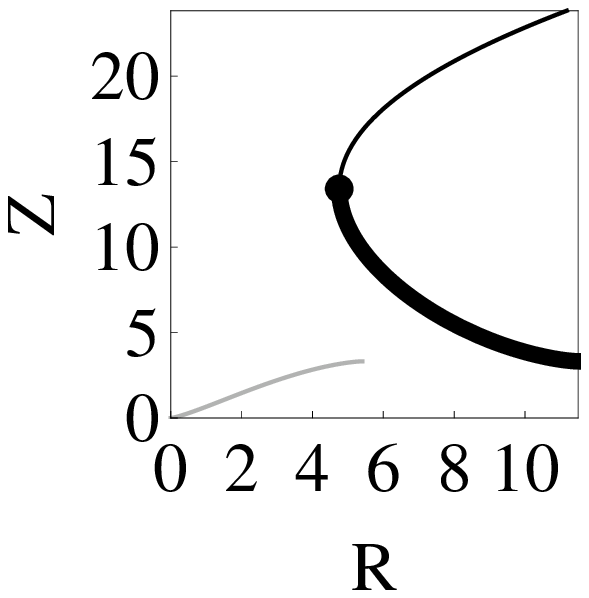}&
                 \includegraphics[clip,height=2.5cm]{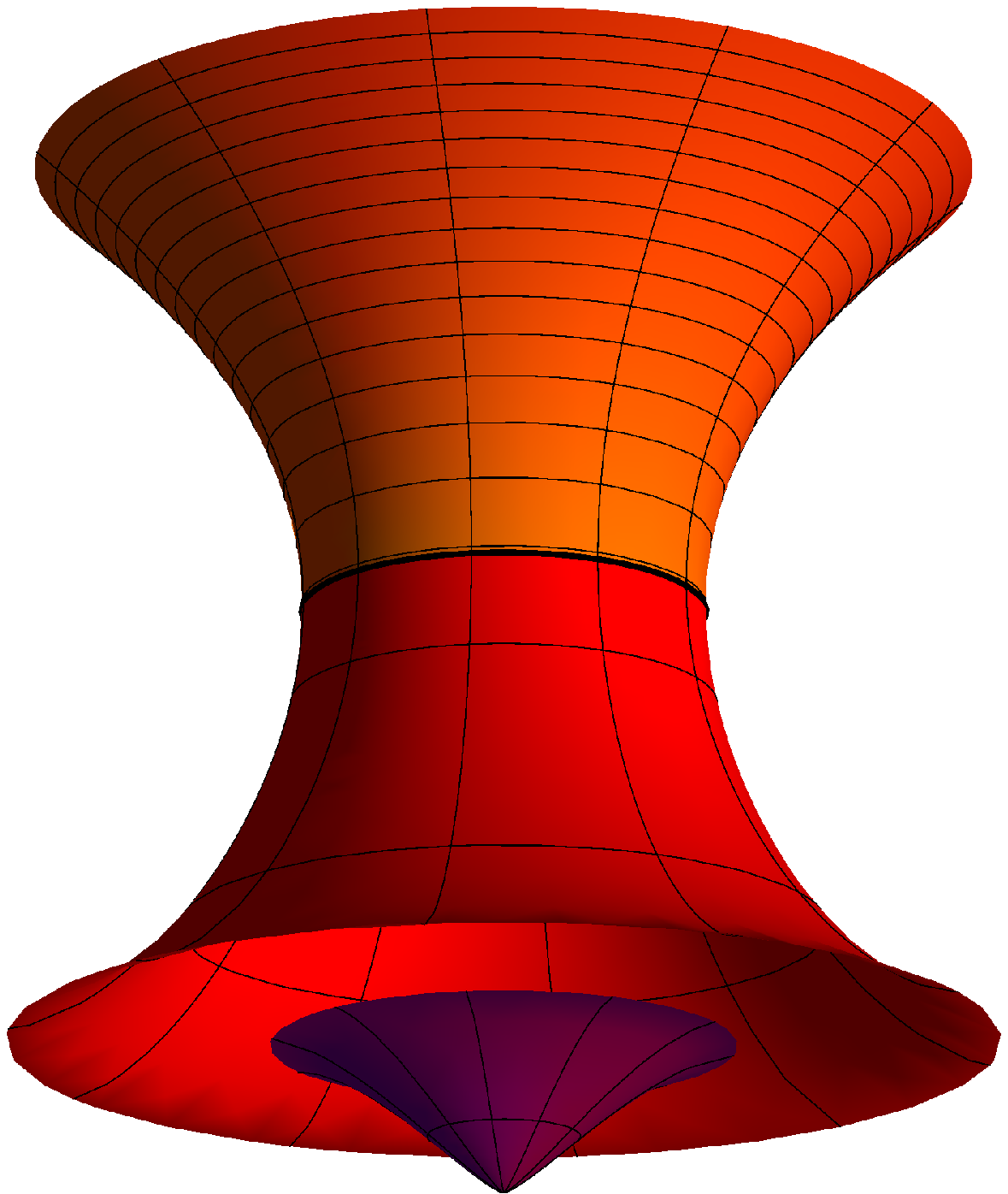}\\
       \includegraphics[clip,height=2.5cm]{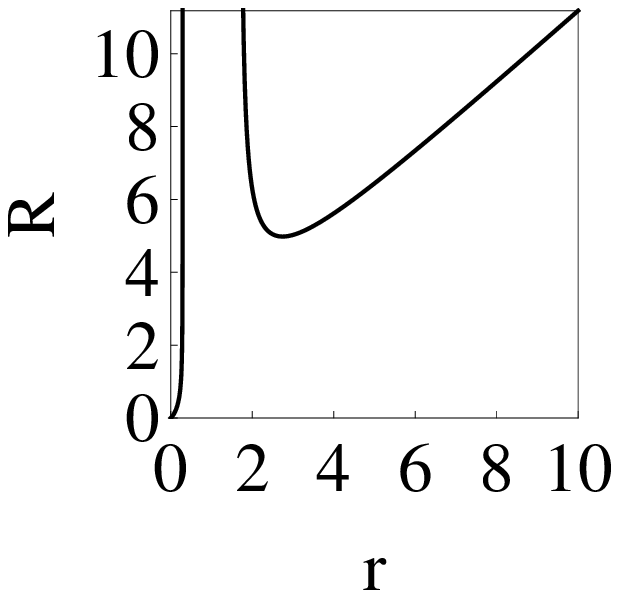}&
        \includegraphics[clip,height=2.5cm]{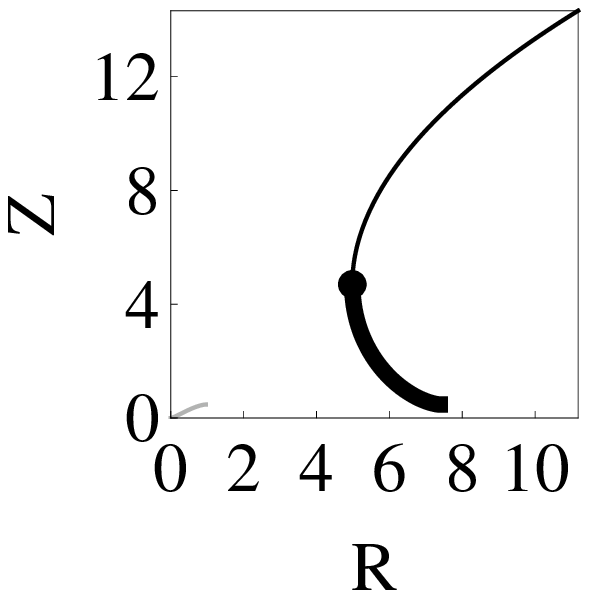}&
                 \includegraphics[clip,height=2.5cm]{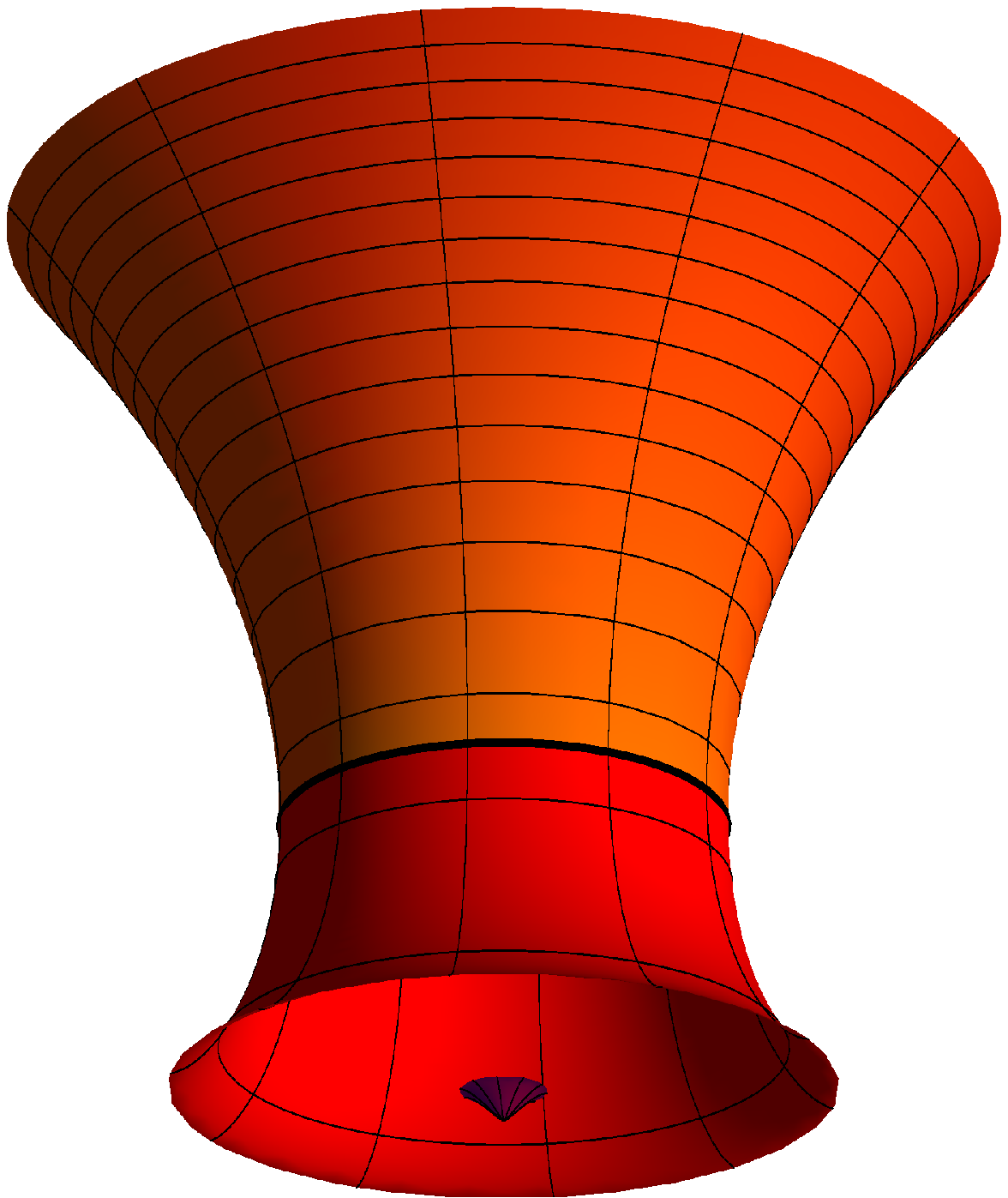}\\
\end{tabular}
    \caption{Class 3 of the embedding diagrams (black hole). The optical
    geometry case is shown. Top: $\omega = 0.55$. Bottom: $\omega = 1.00$.
    The colors and curves have the same meaning as in Fig.~\ref{fig_5}.
    There are two  disconnected ``spaces''
   here: the ``outer'' space outside the outer horizon, and the ``inner'' space
   inside the inner horizon and around the central singularity.
   Note that the embedding diagrams do not end at
   the inner and outer horizons, but at radii corresponding to the embedability limits,
   which are located close to the horizons. Near embedability limits, the embedding diagrams have
   horizontal asymptotes, as $\mathrm{d} Z/\mathrm{d} R = 0$. This cannot be nicely illustrated. Thus, in the Figure,
   the surfaces stop slightly before the embedability limits.}
    \label{fig_6}
    \end{center}
 \end{figure}
%
%
%
%

\section{Discussion}
%

%
\subsection{Comparison between KS and Reissner-Nordstr\"om cases}
\label{KS-and-RN}
As mentioned before, the qualitative properties of the KS solution resemble those of Reissner-Nordstr\"om spacetime.
In particular, we see from Fig. \ref{fig_3b} that the radii corresponding to horizons and photon orbits have the same structure
in both spacetimes, when instead of $\omega$ the horizontal axis displays $q^{-2}$, where
$q=Q/M$ is the charge-to-mass parameter in RN spacetime. The antigravity radius also has a similar behavior in both cases,
meaning that in the
naked-singularity region it is always positive and goes to infinity if the parameter of Fig. \ref{fig_3} or Fig. \ref{fig_3b}
goes to zero, whereas in the black-hole region the static region between the singularity and the inner horizon is of antigravity. The difference appears in the embedding limits for optical (and simply projected) geometry, as first discussed
in \cite{Kri-Son-Abr:1998:GRG:}. As shown in Fig. \ref{fig_3b} for optical geometry,
in the black-hole case the inner region can be embedded only in KS spacetime. Also, the embedability limit in RN
is always below the antigravity radius for the naked-singularity case.

Concerning the simply projected geometry, while in KS spacetime embedability is always possible in static regions,
Eqs. (\ref{geodesic-circumferential-direct}) and (\ref{dZ-dR}) imply
that the embedability condition in RN is given by $r\geq q^2/2$, and since the horizons are given by
$r_\pm=1\pm\sqrt{1-q^2}$ there is, in the black-hole case, a region inner to $r_-$ (the inner horizon) which can be embedded.
However, this region never extends up to the singularity. In the naked-singularity case the embedability limit is always
below the antigravity radius, which is given by $r=q^2$.

%
\subsection{Antigravity and centrifugal force reversal}
\label{antigravity-centrifugal-reversal}
In the original ``physical''  spacetime (\ref{line.element.schwarzscild}), one considers
a~ge\-ne\-ral circular motion, with the four-velocity,
\begin{equation}
\frac{\mathrm{d} x^i}{\mathrm{d} s} = u^i = \gamma \left( N^i + v \tau^i \right),
\label{four-velocity}
\end{equation}
where $N^i$ is a timelike unit vector in the $t$ direction, and $\tau^i$ is a spacelike unit
vector in the $\phi$ direction\footnote{$N^i$ is parallel to the Killing vector $\eta^i =
\delta^i_{~t}$ that exists due to time symmetry, and $\tau^i$ is parallel to the Killing
vector $\xi^i = \delta^i_{~\phi}$ that exists due to axial symmetry. Thus, $N^i$ and
$\tau^i$ are covariantly defined, and therefore $v$ and $\gamma = (1 - v^2)^{-1/2}$ are also
covariantly defined.}. We will define the orbital velocity $V =\gamma v$. The acceleration
in circular motion equals
\begin{equation}
a_i = u^k \nabla_k u_i = g_i + c_i,
\label{four-acceleration}
\end{equation}
where the ``gravitational'' and ``centrifugal'' accelerations are given by formulae which
refer (formally) to the optical space,
\begin{eqnarray}
{\rm gravitational} && \qquad g_i = N^k \nabla_k N_i = \nabla_i \Phi, \qquad \Phi = -\frac{1}{2}\ln f,
\label{gravitational} \\
{\rm centrifugal} && \qquad c_i = \frac{\,\,V^2}{\cal R}\,\lambda_i, \qquad \lambda_i = \frac{\nabla_i
{\tilde r}}{[(\nabla_k {\tilde r})(\nabla^k {\tilde r})]^{1/2}}\,.
\label{centrifugal}
\end{eqnarray}
We stress that $g_i$ and $c_i$ ``live'' in the original ``physical'' spacetime, and that the
reference to the optical space is made here mostly for a formal convenience.

Not {\it only} for convenience, however. This reference also helps to strengthen a
physical intuition, as we will see later when discussing the embedding diagrams of the KS
optical space.

In the above formulae (\ref{gravitational})-(\ref{centrifugal}), the ``gravitational
potential'' $\Phi$ is given by the optical space conformal factor (\ref{conformal-metric}),
and both the curvature radius of the orbit ${\cal R}$ and the unit space-like vector
orthogonal to the trajectory $\lambda_i$ are defined (and calculated) in the optical space.

For the {\it free, geodesic} circular orbits it should be $a_i = 0$. Therefore a necessary
condition for the existence of the geodesic orbit is that the gravitational and centrifugal
accelerations point to the {\it opposite} radial directions,
\begin{equation}
g_i c^i < 0
\label{gravity-centrifugal-opposite}
\end{equation}
in the optical space. In a situation familiar from Newtonian physics, gravity always
pulls inwards and the centrifugal force always pushes out, and therefore the condition
(\ref{gravity-centrifugal-opposite}). In KS spacetimes the situation is more complex, as
two rather unusual effects are present: antigravity and  centrifugal force reversal.

Let us define the ``signs'' of gravitational and centrifugal accelerations by
\begin{equation}
[g] = {\rm sign}(g^i k_i), ~~[c] = {\rm sign}(c^i k_i), \qquad {\rm where}~~ k_i =\nabla_i r.
\label{sign-gravity-centrifugal}
\end{equation}
It should be obvious that gravity pulls in, pushes out or vanishes depending on the sign
of $[g]$,
\begin{equation}
\hskip-0.01truecm [g] = \left\{
    \begin{array}{ll}
        +1  & \mbox{gravity pushes out (``antigravity'')}\\
        ~~0 & \mbox{gravity vanishes} \\
        -1  & \mbox{gravity pulls in }
    \end{array}
\right.
\label{change-sign-gravity}
\end{equation}
and that the same is true for the centrifugal force,
\begin{equation}
\hskip-0.01truecm [c] = \left\{
    \begin{array}{ll}
        +1  & \mbox{centrifugal force  pushes out}\\
        ~~0 & \mbox{centrifugal force vanishes}    \\
        -1  & \mbox{centrifugal force pulls in (``reversal'')}
    \end{array}
\right.
\label{change-sign-centrifugal}
\end{equation}

\begin{figure}[htbp]
\begin{center}
 \begin{tabular}{cc}
\includegraphics[clip,height=2.3cm]{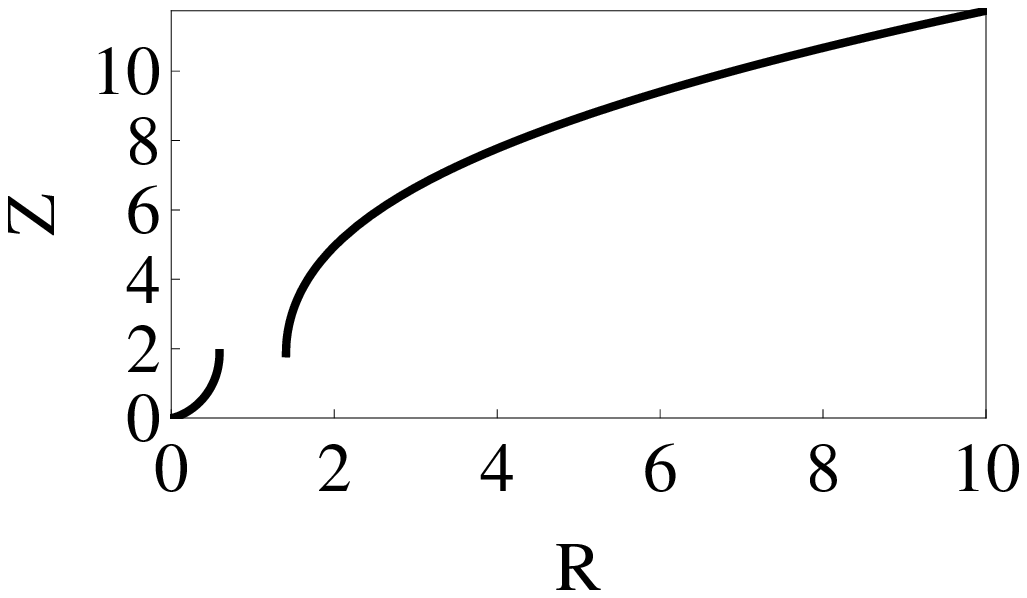}&
\includegraphics[clip,height=2.3cm]{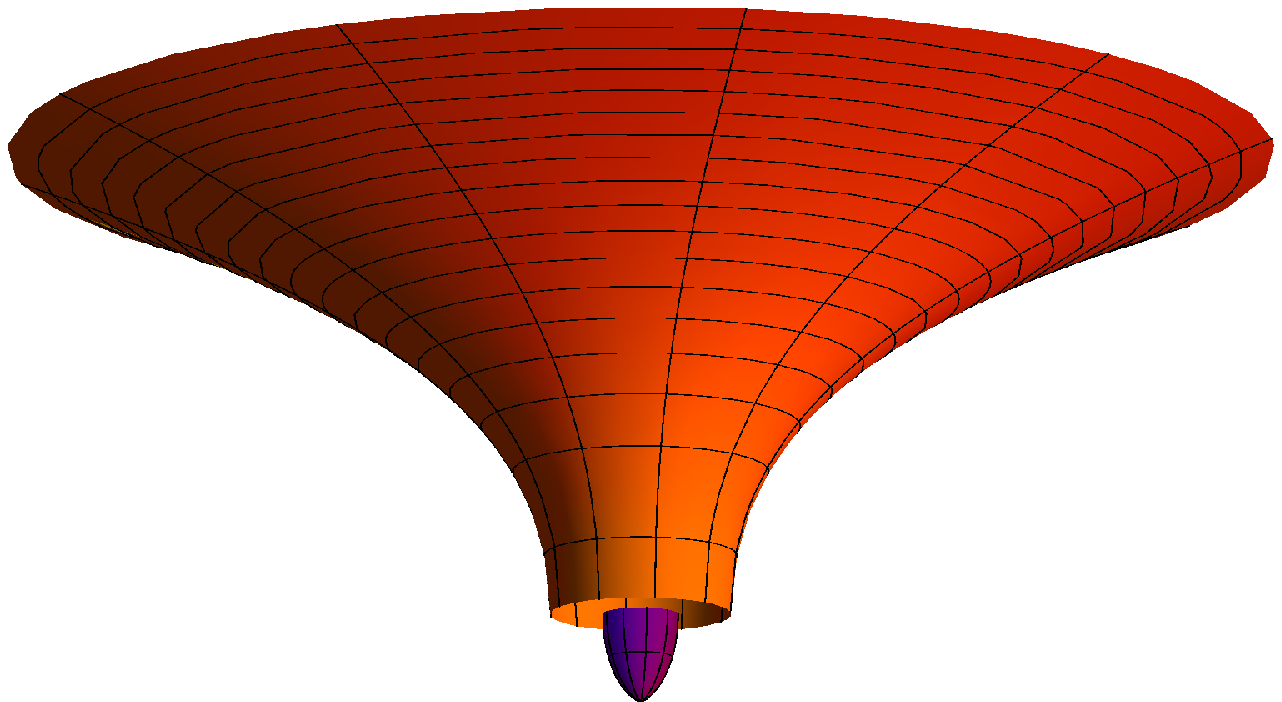}\\
\multicolumn{2}{c}{$\omega = 0.600$} \\ \hline
\includegraphics[clip,height=2.3cm]{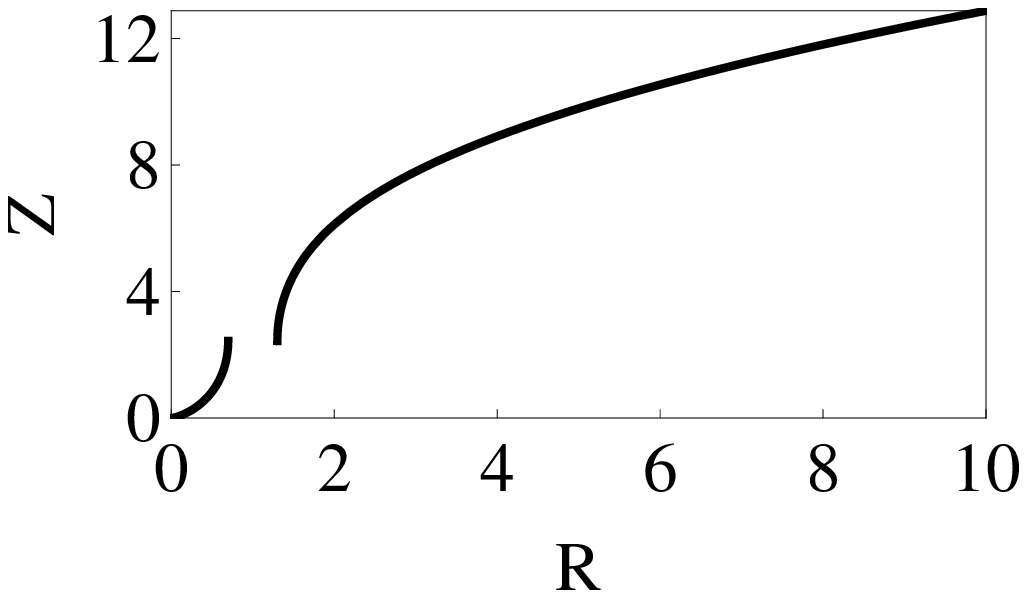}&
\includegraphics[clip,height=2.3cm]{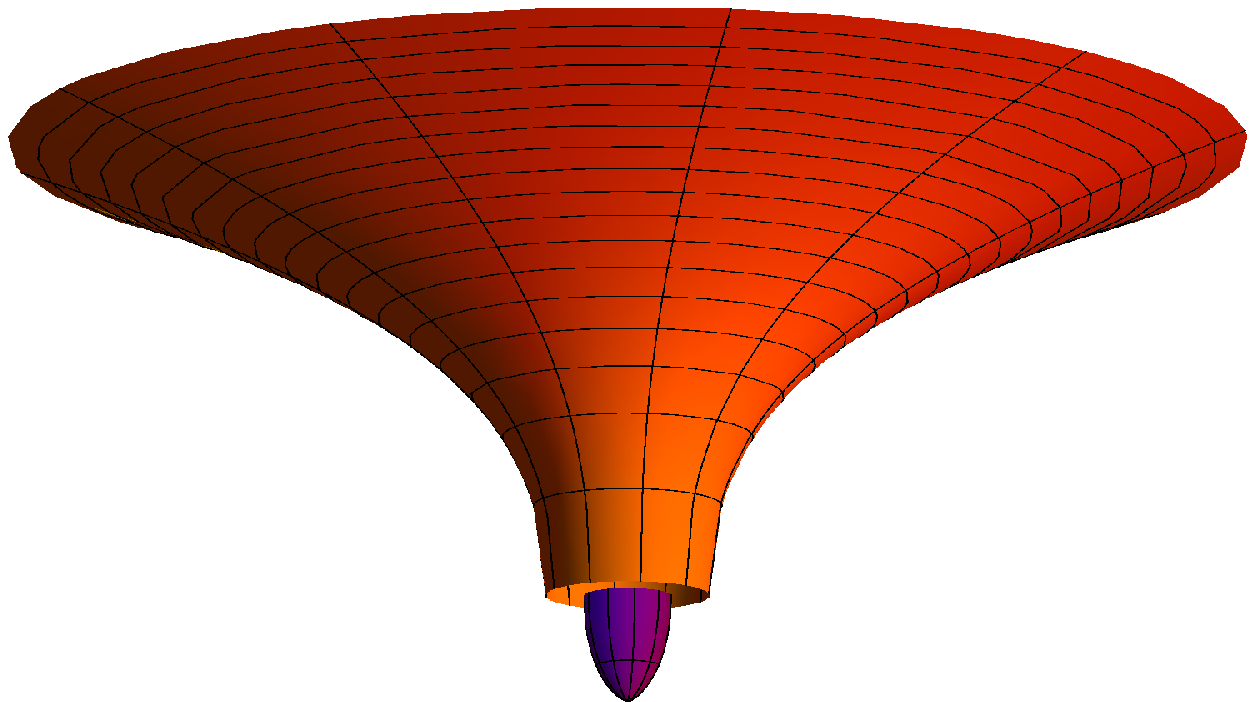}\\
\multicolumn{2}{c}{$\omega = 0.550$} \\ \hline
\includegraphics[clip,height=2.3cm]{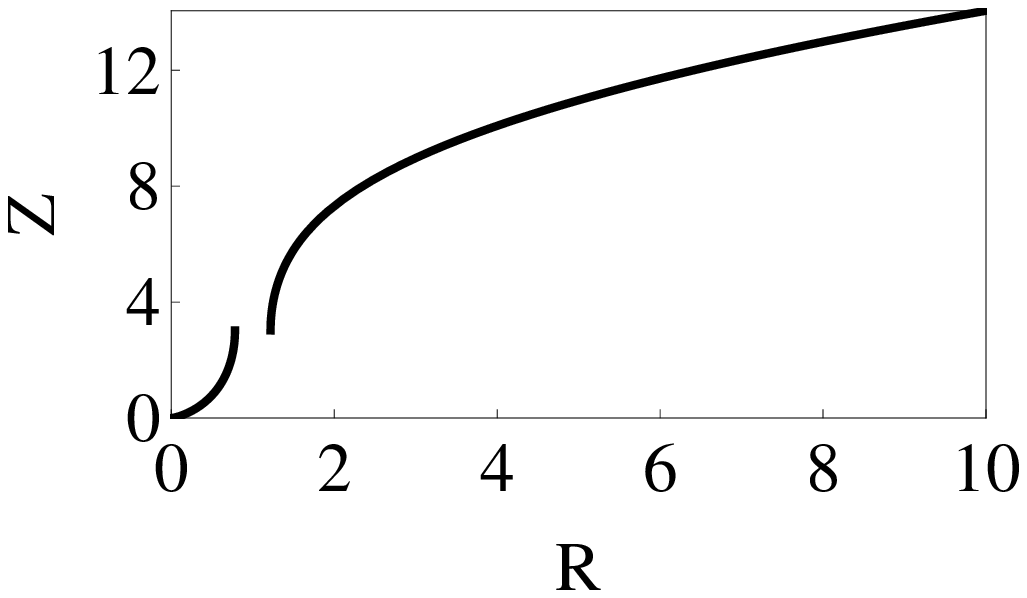}&
\includegraphics[clip,height=2.3cm]{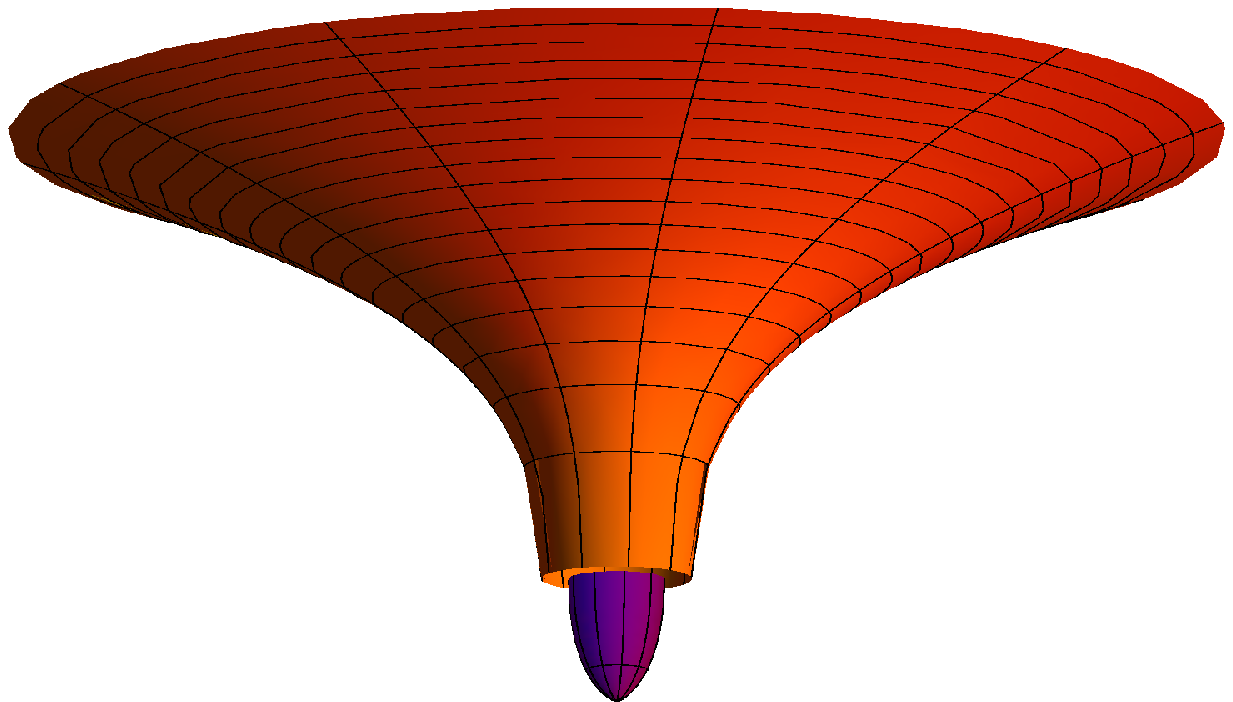}\\
\multicolumn{2}{c}{$\omega = 0.525$} \\ \hline
\includegraphics[clip,height=2.3cm]{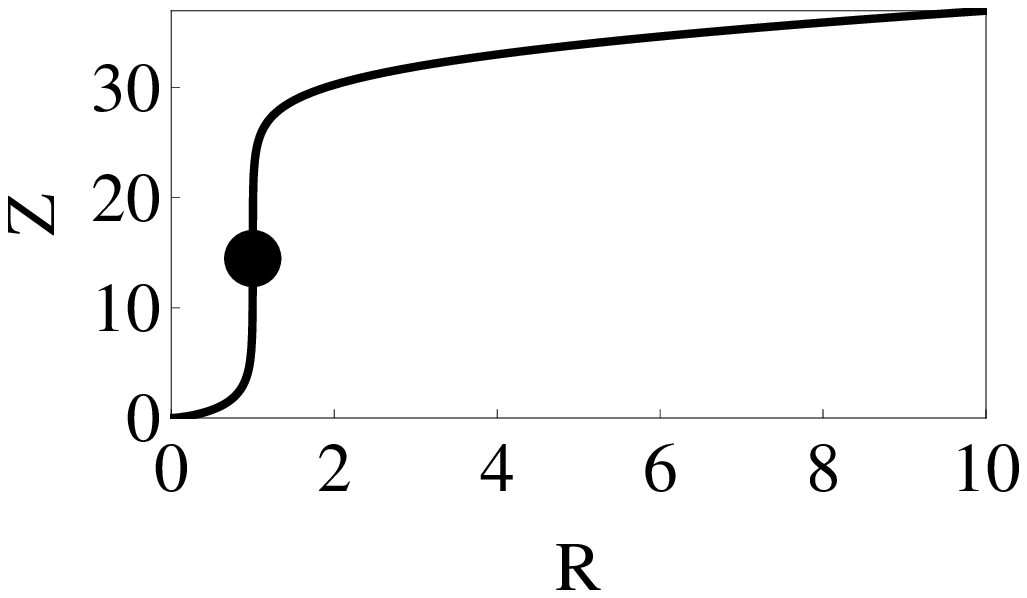}&
\includegraphics[clip,height=2.3cm]{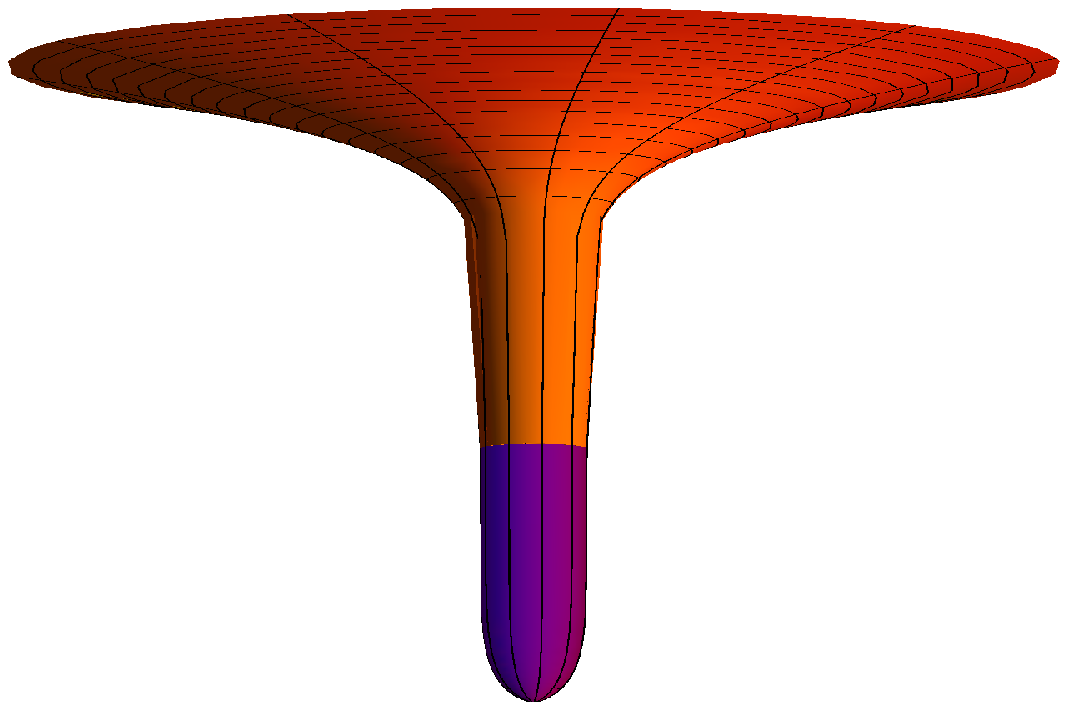}\\
\multicolumn{2}{c}{$\omega = 0.500$} \\ \hline
\includegraphics[clip,height=2.3cm]{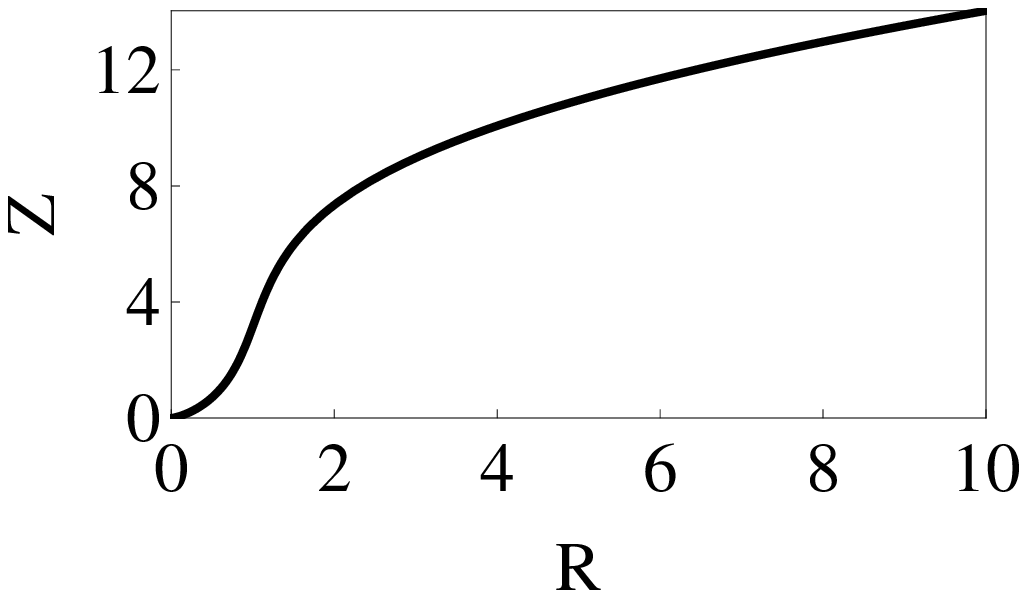}&
\includegraphics[clip,height=2.3cm]{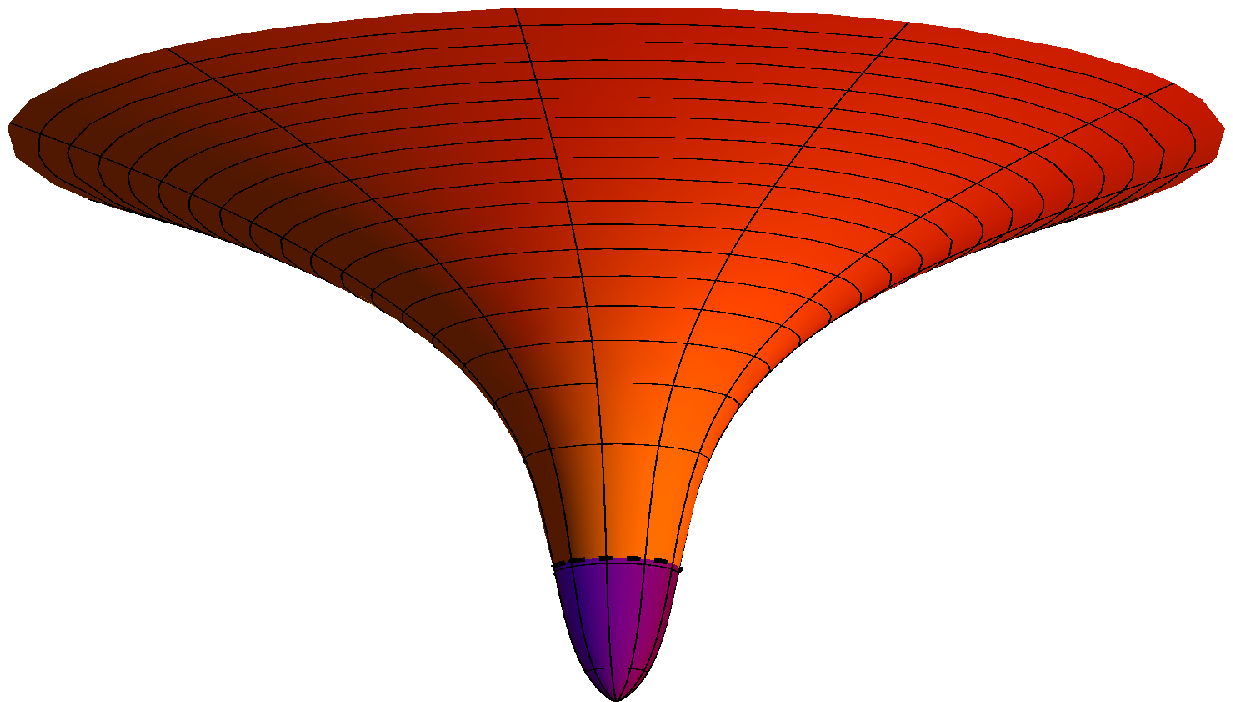}\\
\multicolumn{2}{c}{$\omega = 0.475$} \\ \hline
\includegraphics[clip,height=2.3cm]{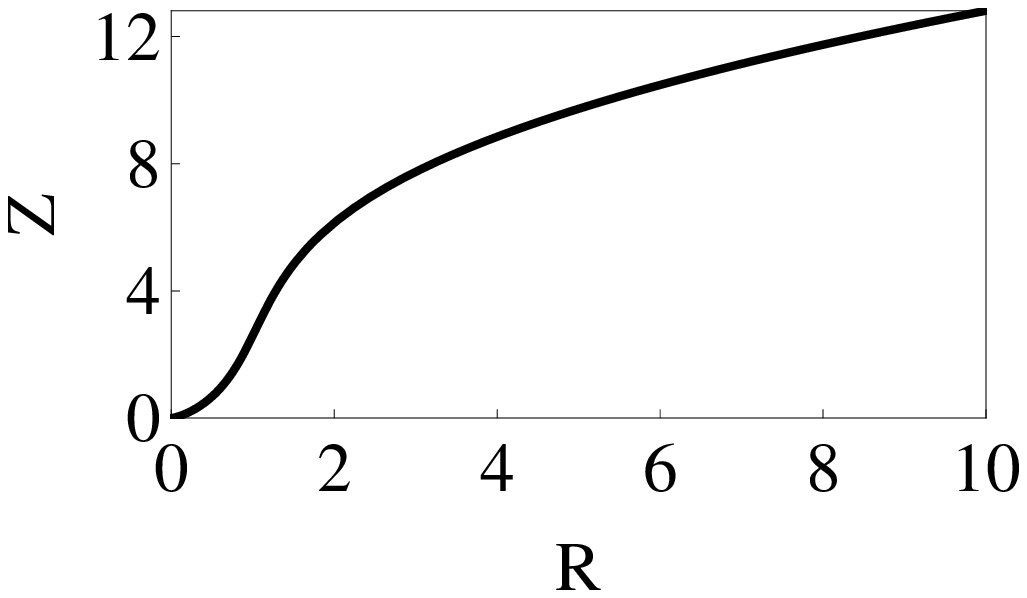}&
\includegraphics[clip,height=2.3cm]{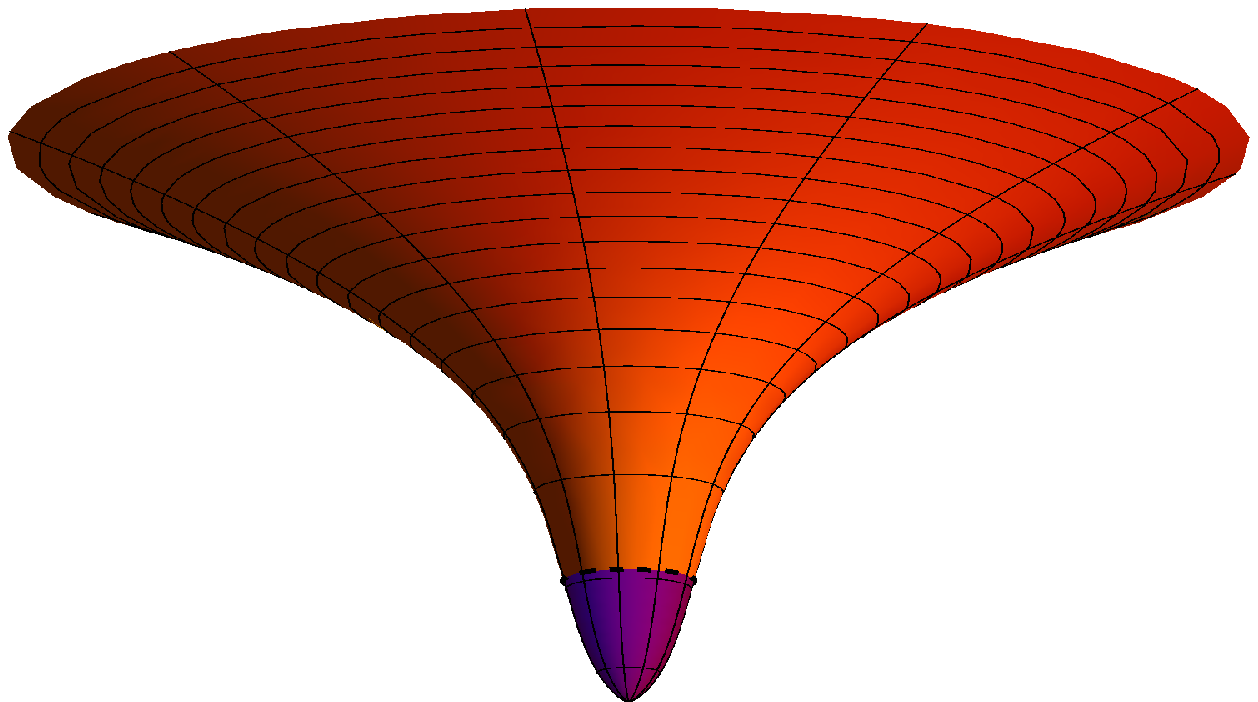}\\
\multicolumn{2}{c}{$\omega = 0.450$} \\ \hline
\includegraphics[clip,height=2.3cm]{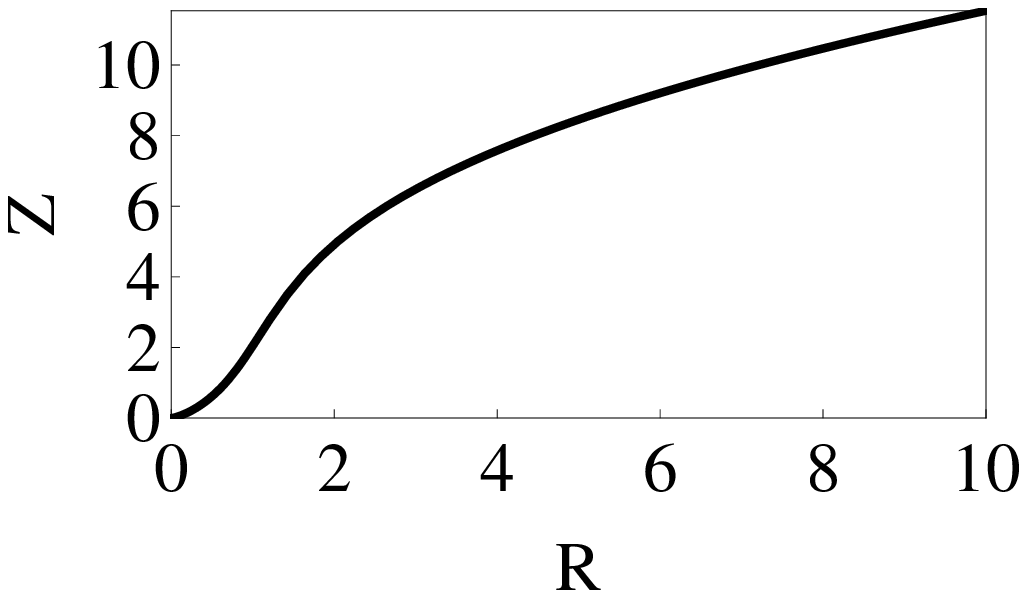}&
\includegraphics[clip,height=2.3cm]{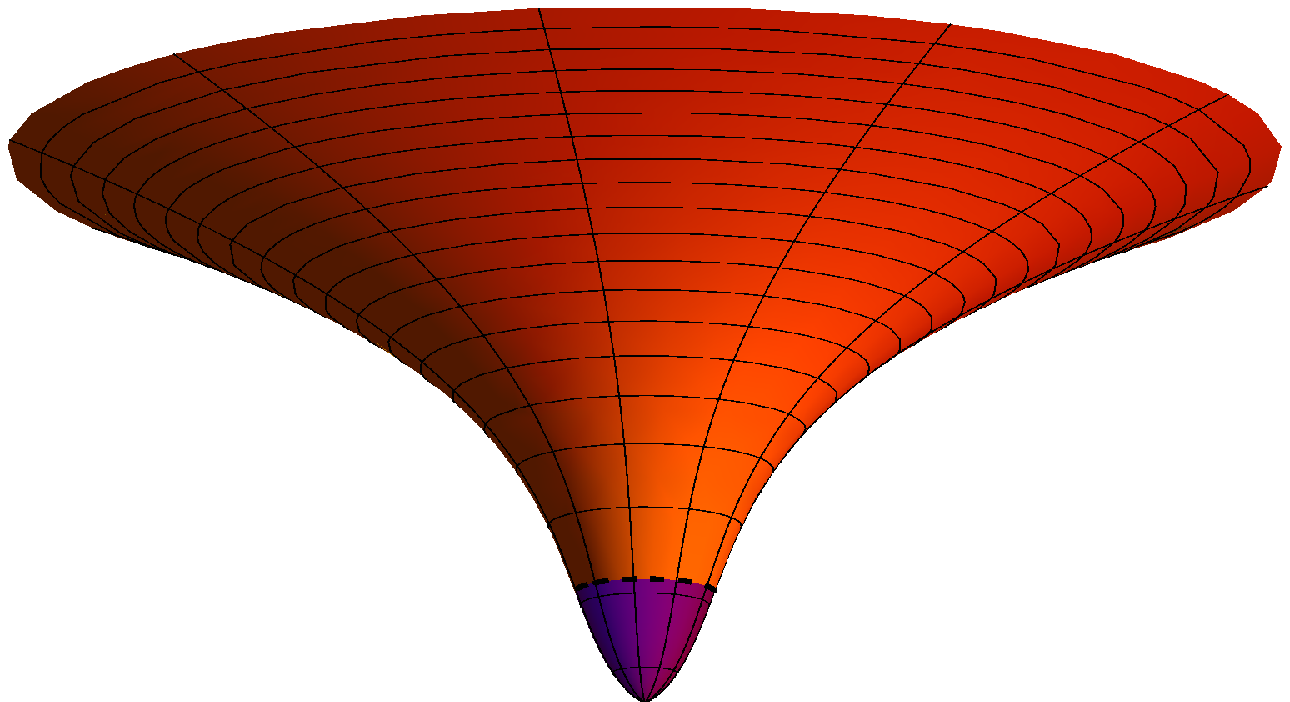}\\
\multicolumn{2}{c}{$\omega = 0.400$} \\ \hline
\end{tabular}
\caption{A sequence of KS objects in simply projected space, showing a transition between the naked singularity
and the black hole state (from bottom to top along the sequence shown in the Figure). The
transition is a discontinuous one. For $\omega<1/2$ there is no horizon, and at $\omega =
1/2$ a finite-size horizon is formed. Its location is shown by a black dot and a solid line
in the $\omega=0.500$ panel.}
\label{fig_8}
\end{center}
\end{figure}

In the embedding diagrams, the region of antigravity, close to the naked singularity at
$r=0$ is represented by a strong violet color, and the sphere (a circle on the equatorial plane)
of zero gravity is shown in the embedding profiles by light grey. In the region of
antigravity the centrifugal force also pushes out, so condition
(\ref{gravity-centrifugal-opposite}) is not fulfilled---free circular (geodesic)
orbits are not possible there. Also note that, in the black-hole case, the whole region
between the singularity and the inner horizon is an antigravity region---free circular
(geodesic) orbits are not possible there.

The centrifugal force reversal happens in the region where, in optical geometry, the
circumferential radius ${\tilde r} = R$ decreases in the direction of increasing $r_*$,
colored red in Figs.~\ref{fig_5}, \ref{fig_6}. In
these regions the gravitational force pulls in, so condition
(\ref{gravity-centrifugal-opposite}) is not fulfilled --- free circular orbits are not
possible there. The centrifugal force vanishes (for any finite value of the orbital velocity
$V$) at the {\it turning points} where ${\tilde r} = R$ has a local extremum (minimum or
maximum). The minima corresponds to unstable, and maxima to stable circular photon orbits.
These orbits are distinguished by solid circles in the optical geometry embedding diagrams.
%
%

%
%
\section{Conclusions}
%
We have discussed geometrical properties of KS spacetimes in terms of embedding
diagrams.
The KS metric describes a static, spherically symmetric spacetime in the Ho{\v r}ava quantum
gravity theory. The non-Einsteinian (quantum) features of the metric are governed by a
single parameter $\omega$, which has dimension [cm]$^{-2}$. The central KS object, with mass
$M$ corresponds to either a black hole (when $\omega M^2 > 1/2$) or a naked singularity
(when $\omega M^2 < 1/2$).

We show that a (hypothetical) transition from a naked singularity
to a black hole state may be discontinuous, as clearly demonstrated by a sequence of KS spacetimes (Fig. \ref{fig_8}),
 with the horizon size starting from a finite
value. This may have an observable astrophysical signature.
A similar discontinuity may occur also in the case of the conversion of a Kerr naked singularity into a near-extreme Kerr black 
hole~\cite{Stu:1980:Bull.Astr.Inst.Czechoslov:}.

We also discussed the phenomena of antigravity and of centrifugal
force reversal present in the KS spacetimes. The first one is present
for naked singularities, but not outside black holes,
the latter  for both naked singularities and black holes.

\begin{acknowledgements}
{The~authors 
acknowledge the~ Research centre of theoretical physics and astrophysics, Faculty of Philosophy and Science, Silesian Univerzity in Opava. MAA and WK acknowledge the Polish NCN grants UMO-2011/01/B/ST9/05439 and 2013/10/M/ST9/00729. KG would like to express her gratitude to the~internal grants of the~Silesian University in Opava FPF SGS/11,23/2013. ZS would like to thank Albert Einstein Center for Gravitation and Astrophysics supported by the Czech Science Foundation grand No. 14-37086G.  RSSV
acknowledges the financial support of Funda\c{c}\~ao de Amparo \`a Pesquisa do Estado de
S\~ao Paulo (FAPESP), Grants 2010/00487-9, 2013/01001-0 and 2015/10577-9.}
\end{acknowledgements}


%
\clearpage

\appendix
\normalsize
\section{Reference guide to Ho{\v r}ava's gravity}
%
Here we give some additional references that may be useful in the context of our article.
\par                                                                
The Ho\v{r}ava gravity (or ``Ho\v{r}ava-Lifshitz'' gravity as it is often called) is
considered as one of the most promising approaches to the quantum gravity.
(see \cite{And-etal:2012:PHYSR4:,Gri-Hor-MelT:2012:JHEP:,Gri-Hor-MelT:2013:PHYSRL:,Hor:2009:PHYSR4:,Hor:2009:PHYSRL:,Hor-MelT:2010:PHYSR4:,Lib-Mac-Sot:2012:PHYSRL:,Ver-Sot:2012:PHYSR4:}).
\par                                                                
The solutions of the Ho\v{r}ava-Lifshitz effective gravitational equations have been found
in
\cite{Bar-Sot:2012:PHYSRL:,Bar-Sot:2013:PHYSRL:,Bar-Sot:2013:PHYSR4:}.
\par                                                                
The spherically symmetric solution of the Schwarz\-schild character has been found in the
framework of the modified Ho\v{r}ava model and are given by the Kehagias-Sfetsos (KS)
metric, which allows for existence of both black hole and naked singularity spacetimes.
(see \cite{Keh-Sfe:2009:PhysLetB:,Mu-In:2009:JHEP:}).
\par                                                                
In connection to the accretion phenomena, the KS metric has been extensively studied in a
series of works related both to the particle motion and optical phenomena (the last two
references below) that can be relevant for tests of validity of the Ho\v{r}ava-Lifshitz
gravity. Here we demonstrate the properties of the KS spacetime using the embedding diagrams
of both the direct geometry and the optical geometry reflecting some hidden properties of
the spacetime and its geodetic structure.
(see \cite{Abd-Ahm-Hak:2011:PHYSR4:,Ali-Sen:2010:PHYSR4:,Ata-Abd-Ahm:2013:AstrSpaScie:,Eno-etal:2011:PHYSR4:,Hak-etal:2010:MPLA:,Hak-Abd-Ahm:2013:PHYSR4:,Hor-Ger-Ker:2011:PHYSR4:}).
\par                                                                
In this paper we discussed not only the KS black holes, but also the KS naked singularities.
Some phenomena found in the KS naked singularities are similar to these found for the
Reissner-Nordstr{\"o}m naked singularities,
(e.g \cite{Stu-Hed:2002:APS:}),
the braneworld naked singularities
(e.g \cite{Ali-etal:2013:CLAQG:,Stu-Kol:2012:JCAP:,Stu-Kot:2009:GenRelGrav:})and the Kerr naked singularities
(e.g \cite{Pat-Jos:2011:CLAQG:,Sche-Stu:2013:JCAP:,Stu-Sche:2010:CLAQG:,Stu-Hle-Tru:2011:CLAQG:}).

%
%

%


\end{document}